\documentclass[aap,MSNbibl,seceqn,citesort,dvips]{arximspdf}
\usepackage{mathrsfs}

\doi{10.1214/11-AAP776}
\volume{22}
\issue{1}
\pubyear{2012}
\firstpage{363}
\lastpage{406}

\makeatletter

\newtheorem{Thm}{Theorem}[section]
\newtheorem{Prop}[Thm]{Proposition}
\newtheorem{Lemma}[Thm]{Lemma}
\newtheorem{Cor}[Thm]{Corollary}
\newproclaim{Def}[Thm]{Definition}
\newproclaim{Remark}[Thm]{Remark}
\newproclaim{Example}[Thm]{Example}
\newproclaim{Assumptions}[Thm]{Assumptions}

\newcommand{\eps}{\varepsilon}
\newcommand{\R}{\mathbb{R}}
\newcommand{\F}{\mathbb{F}}
\newcommand{\cF}{\mathcal{F}}
\newcommand{\cP}{\mathcal{P}}
\newcommand{\cA}{\mathcal{A}}
\newcommand{\cE}{\mathcal{E}}
\newcommand{\sC}{\mathscr{C}}
\newcommand{\sD}{\mathscr{D}}
\newcommand{\sM}{\mathscr{M}}
\newcommand{\sN}{\mathscr{N}}

\newcommand{\tR}{{\widetilde{R}}}
\newcommand{\tsC}{{\widetilde{\mathscr{C}}}}
\newcommand{\tg}{{\tilde{g}}}

\newcommand{\sgn}{\operatorname{sign}}
\newcommand{\Var}{\operatorname{Var}}
\newcommand{\esssup}{\mathop{\operatorname{ess}\sup}}
\newcommand{\argmax}{\mathop{\arg\max}}

\newcommand{\hX}{\widehat{X}}
\newcommand{\hY}{\widehat{Y}}

\newcommand{\tc}{\tilde{c}}
\newcommand{\hc}{\hat{c}}
\newcommand{\cpi}{\check{\pi}}
\newcommand{\cc}{\check{c}}
\newcommand{\cy}{\check{y}}
\newcommand{\hpi}{\hat{\pi}}
\newcommand{\hkappa}{\hat{\kappa}}
\newcommand{\ckappa}{\check{\kappa}}
\newcommand{\cX}{\check{X}}
\newcommand{\tpi}{\tilde{\pi}}
\newcommand{\tZ}{\widetilde{Z}}
\newcommand{\sint}{\stackrel{\mbox{\tiny$\bullet$}}{}}

\makeatother

\begin{document}
\begin{frontmatter}

\title{The Bellman equation for power utility maximization with
semimartingales}
\runtitle{Bellman equation for power utility}

\begin{aug}
\author[A]{\fnms{Marcel} \snm{Nutz}\corref{}\thanksref{t1}\ead[label=e1]{mnutz@math.columbia.edu}}
\runauthor{M. Nutz}
\affiliation{ETH Z\"{u}rich}
\address[A]{Department of Mathematics\\
Columbia University\\
2990 Broadway\\
New York, New York 10027\\
USA\\
\printead{e1}} 
\end{aug}

\thankstext{t1}{Supported by Swiss National Science Foundation Grant
PDFM2-120424/1.}

\received{\smonth{12} \syear{2009}}
\revised{\smonth{3} \syear{2011}}

%
\begin{abstract}
We study utility maximization for power utility random fields with and
without intermediate consumption in a general semimartingale model with
closed portfolio constraints. We show that any optimal strategy leads
to a solution of the corresponding Bellman equation. The optimal
strategies are described pointwise in terms of the opportunity process,
which is characterized as the minimal solution of the Bellman equation.
We also give verification theorems for this equation.
\end{abstract}

%

\end{frontmatter}

\section{Introduction}\label{intro}

A classical problem of mathematical finance is the maximization of
expected utility obtained from consumption or from terminal wealth.
This paper focuses on power utility functions and presents the
corresponding dynamic programming in a general constrained
semimartingale framework. The homogeneity of these utility functions
leads to a factorization of the value process into a part depending on
the current wealth and the so-called opportunity process~$L$.
In our setting, the Bellman equation describes the drift rate of $L$
and clarifies the local structure of our problem.
Finding an optimal\vspace*{1pt} strategy boils down to
maximizing a random function $y\mapsto g(\omega,t,y)$ on $\R^d$ for
every state $\omega$ and date $t$. This function is
given in terms of the semimartingale characteristics of $L$ as well as
the asset returns, and its maximum yields the drift rate of $L$. The
role of the opportunity process is to augment the information contained
in the return characteristics in order to have
a local sufficient statistic for the global optimization problem.

We present three main results. First, we show that if there exists an
optimal strategy for the utility maximization problem, \textit{the
opportunity process~$L$ solves the Bellman equation} and we provide a
local description of the optimal strategies.
We state the Bellman equation in two forms, as an identity for the
drift rate of $L$ and as a backward stochastic differential equation
(BSDE) for $L$.
Second, we characterize the
opportunity process as the \textit{minimal solution} of this equation.
Finally, given some solution and
an associated strategy, one can ask whether the strategy is optimal and
the solution is the opportunity process.
We present two different approaches which lead to \textit{verification
theorems} not comparable in strength unless
the constraints are convex.

The present dynamic programming approach should be seen as
complementary to convex duality, which remains the only method to
obtain \textit{existence} of optimal strategies in general models (see
Kramkov and Schachermayer \cite{KramkovSchachermayer99}, Karatzas and
\v
{Z}itkovi\'c \cite{KaratzasZitkovic03}, Karatzas and Kardaras \cite
{KaratzasKardaras07}). However, convex duality alone offers limited
insight into the optimal strategies for incomplete markets.
In some cases, the Bellman equation can be solved directly by analytic
methods, for example, in the setting of Example \ref
{exHuImkellerMuller} with continuous asset prices or in the L\'evy
process setting of Nutz \cite{Nutz09c}. In addition to the existence,
one then obtains a way to compute the optimal strategies (at least
numerically) and study their properties.

This paper is organized as follows.
The next section specifies the optimization problem in detail, recalls
the opportunity process and the martingale optimality principle and
fixes the notation for the characteristics. We also introduce
set-valued processes describing the
budget condition and state the assumptions on the portfolio constraints.
Section \ref{sebellmanEqn} derives the Bellman equation, first as a
drift condition and then as a BSDE.
It becomes more explicit as we specialize to the case of continuous
asset prices.
The definition of a solution of the Bellman equation is given in
Section \ref{seminimality}, where we show
the minimality of the opportunity process.
Section~\ref{severification} deals with the verification problem, which
is converse to the derivation of the Bellman equation since it requires
the passage from the local maximization to the global optimization problem.
We present an approach via the value process and a second approach via
a deflator, which corresponds to the dual problem in a suitable setting.
Appendix \ref{seMeasSelection} belongs to Section \ref{sebellmanEqn}
and contains the measurable selections for the construction of the
Bellman equation. It is complemented by Appendix \ref
{setransformedModel}, where we construct an alternative parametrization
of the market model
by representative portfolios.

\section{Preliminaries}

The following notation is used. If $x,y\in\R$, we denote $x^+=\max\{
x,0\}$ and $x\wedge y=\min\{x,y\}$. We set $1/0:=\infty$ where
necessary. If $z\in\R^d$ is a $d$-dimensional vector, $z^i$ is its
$i$th coordinate, $z^\top$ its transpose and
$|z|=(z^\top z)^{1/2}$ the Euclidean norm. If $X$ is an $\R^d$-valued
semimartingale
and $\pi$ is an $\R^d$-valued predictable integrand, the vector
stochastic integral is a scalar semimartingale with initial value zero
and denoted by~$\int\pi\,dX$ or by~$\pi\sint X$. The quadratic
variation is the $d\times d$-matrix $[X]:=[X,X]$ and
if $Y$ is a scalar semimartingale, $[X,Y]$ is the $d$-vector with
$[X,Y]^i:=[X^i,Y]$. When the reference measure is understood, relations
between measurable functions hold almost everywhere unless otherwise
mentioned. Our reference for any unexplained notion from stochastic
calculus is Jacod and Shiryaev \cite{JacodShiryaev03}.

\subsection{The optimization problem}

We fix the time horizon $T\in(0,\infty)$ and a stochastic basis
$(\Omega
,\cF,\F,P)$, where the filtration
$\F=(\cF_t)_{t\in[0,T]}$ satisfies the usual assumptions of right
continuity and completeness as well as $\cF_0=\{\varnothing,\Omega\}
$ $P$-a.s.
We consider an $\R^d$-valued c\`adl\`ag semimartingale $R$ with $R_0=0$
representing the returns of $d$ risky assets.
Their discounted prices are given by the stochastic exponential $S=\cE
(R)=(\cE(R^1),\ldots,\cE(R^d))$; in the financial application, the
components of $S$ are assumed to be positive.
Our agent also has a bank account at his disposal; it does not pay interest.

The agent is endowed with a deterministic initial capital $x_0>0$. A
\textit{trading strategy} is a predictable $R$-integrable $\R^d$-valued
process $\pi$, where $\pi^i$ indicates the fraction of wealth (or the
portfolio proportion) invested in the $i$th risky asset. A~\textit
{consumption strategy} is a nonnegative optional process $c$ such that
$\int_0^Tc_t\,dt<\infty$ \mbox{$P$-a.s.}
We want to consider two cases. Either consumption occurs only at the
terminal time $T$ (utility from ``terminal wealth'' only) or there is
intermediate consumption plus a bulk consumption at the time horizon.
To unify the notation, we introduce the measure $\mu$ on $[0,T]$ by
\[
\mu(dt):=
\cases{
0, &\quad in the case without intermediate consumption, \cr
dt, &\quad in the case with intermediate consumption.}
\]
Let also $\mu^\circ:=\mu+ \delta_{\{T\}}$, where $\delta_{\{T\}}$ is
the unit Dirac measure at $T$.
The \textit{wealth process} $X(\pi,c)$ corresponding to a pair $(\pi,c)$
is defined by the equation
\[
X_t(\pi,c)=x_0+\int_0^t X_{s-}(\pi,c) \pi_s\,dR_s-\int_0^t c_s \mu
(ds),\qquad 0\leq t\leq T.
\]
We define the set of trading and consumption pairs
\[
\cA^0(x_0):=\{(\pi,c)\dvtx X(\pi,c)>0, X_-(\pi,c)>0 \mbox{ and }
c_T=X_T(\pi,c)\}.
\]
These are the strategies that satisfy the budget constraint. The
convention $c_T=X_T(\pi,c)$ means that all the remaining wealth is
consumed at time~$T$. We consider also exogenous constraints imposed on
the agent.
For each $(\omega,t)\in\Omega\times[0,T]$ we are given a set $\sC
_t(\omega)\subseteq\R^d$ which contains the origin.
The set of (constrained) \textit{admissible} strategies is
\[
\cA(x_0):=\{(\pi,c)\in\cA^0(x_0)\dvtx\pi_t(\omega)\in\sC
_t(\omega)
\mbox{ for all }(\omega,t)\},
\]
which is nonempty as $0\in\sC_t(\omega)$. Further assumptions on the
set-valued mapping $\sC$
will be introduced in Section \ref{seConstraintsDegeneracies}. We fix
the initial capital $x_0$ and usually write $\cA$ for $\cA(x_0)$.
Abusing the notation, we write $c\in\cA$ and call $c$ admissible if
there exists $\pi$ such that $(\pi,c)\in\cA$; an analogous convention
is used for similar expressions.

We will often parametrize the consumption strategies as a fraction of wealth.
Let $(\pi,c)\in\cA$ and $X=X(\pi,c)$. Then
\[
\kappa:=\frac{c}{X}
\]
is called the \textit{propensity to consume} corresponding to $(\pi,c)$.
This yields a one-to-one correspondence between the pairs $(\pi,c)\in
\cA
$ and the pairs $(\pi,\kappa)$ such that $\pi\in\cA$ and $\kappa$
is a
nonnegative optional process satisfying $\int_0^T \kappa_s\,ds<\infty$
$P$-a.s.
and $\kappa_T=1$ (see Nutz \cite{Nutz09a}, Remark 2.1, for details).
We shall abuse the notation and identify a consumption strategy with
the corresponding propensity to consume; for example, we write $(\pi
,\kappa)\in\cA$.
Note that
\[
X(\pi,\kappa)=x_0\cE(\pi\sint R - \kappa\sint\mu).
\]
This simplifies verifying that some pair $(\pi,\kappa)$ is admissible
as $X(\pi,\kappa)>0$
implies $X_-(\pi,\kappa)>0$; cf. \cite{JacodShiryaev03}, II.8a.

The preferences of the agent are modeled by a time-additive random
utility function as follows. Let $D$ be a c\`adl\`ag, adapted, strictly
positive process such that
$E[\int_0^T D_s \mu^\circ(ds)]<\infty$
and fix $p\in(-\infty,0)\cup(0,1)$. We define the power utility
random field
\[
U_t(x):=D_t\frac{1}{p}x^p,\qquad x\in(0,\infty), t\in[0,T].
\]
This is the general form of a \textit{$p$-homogeneous} utility random
field such that a constant consumption yields finite
expected utility.
Interpretations and applications for the process $D$ are discussed
in \cite{Nutz09a}.
We denote by $U^*$ the convex conjugate of $x\mapsto U_t(x)$,
%
%
\begin{equation}\label{eqconvexConjugate}
U_t^*(y)=\sup_{x>0} \{U_t(x)-xy\}=-\frac{1}{q}y^{q}D_t^{\beta};
\end{equation}
here $q:=\frac{p}{p-1}\in(-\infty,0)\cup(0,1)$ is the exponent
conjugate to $p$ and the constant $\beta:=\frac{1}{1-p}>0$
is the\vspace*{1pt} relative risk tolerance of $U$.
Note that we exclude the well-studied logarithmic utility (e.g., Goll
and Kallsen \cite{GollKallsen03})
which corresponds to \mbox{$p=0$}.

The \textit{expected utility} corresponding to a consumption strategy
$c\in\cA$ is
$E[\int_0^T U_t(c_t) \mu^\circ(dt)]$, that is,
either $E[U_T(c_T)]$ or $E[\int_0^T U_t(c_t)\,dt+U_T(c_T)]$.
The (value of the) utility maximization problem is said to be \textit
{finite} if
%
%
\begin{equation}\label{eqPrimalProblemFinite}
u(x_0):=\sup_{c\in\cA(x_0)}E\biggl[\int_0^T U_t(c_t) \mu^\circ(dt)
\biggr]<\infty.
\end{equation}
Note that this condition is void if $p<0$ as then $U<0$.
If (\ref{eqPrimalProblemFinite}) holds, a~strategy $(\pi,c)\in\cA(x_0)$
is called \textit{optimal} if $E[\int_0^T U_t(c_t) \mu^\circ(dt)
]=u(x_0)$.

Finally, we introduce the following sets which are of minor importance
and used only in the case $p<0$:
\begin{eqnarray*}
\cA^f&:=& \biggl\{(\pi,c)\in\cA\dvtx \int_0^T U_t(c_t) \mu^\circ
(dt)>-\infty\biggr\}, \\
\cA^{fE}&:=&\biggl\{(\pi,c)\in\cA\dvtx E\biggl[\int_0^T U_t(c_t)
\mu^\circ(dt)\biggr]>-\infty\biggr\}.
\end{eqnarray*}
Anticipating that (\ref{eqPrimalProblemFinite}) will be in force, the
indices stand for ``finite'' and ``finite expectation.'' Clearly $\cA
^{fE}\subseteq\cA^f\subseteq\cA$, and equality holds if $p\in(0,1)$.

%
\subsection{Opportunity process}\label{seoppProc}
We recall the opportunity process, a reduced form of the value process
in the language of control theory.
We assume (\ref{eqPrimalProblemFinite}) in this section, which ensures
that the following process is finite. By \cite{Nutz09a},
Proposition~3.1 and Remark 3.7, there exists a unique c\`adl\`ag
semimartingale~$L$, called \textit{opportunity process}, such that
%
%
\begin{equation}\label{eqOppProcIndep}
L_t \frac{1}{p}(X_t(\pi,c))^p= \mathop{\esssup}_{\tilde{c}\in
\cA(\pi,c,t)} E\biggl[\int_t^T U_s(\tc_s) \mu^\circ(ds)\Big|\cF_t\biggr]
\end{equation}
for any\vspace*{1pt} $(\pi,c)\in\cA$, where
$\cA(\pi,c,t):=\{(\tilde{\pi},\tilde{c})\in\cA\dvtx(\tilde
{\pi
},\tilde{c})=(\pi,c)\mbox{ on }[0,t]\}$.
We note that $L_T=D_T$ and
that $u(x_0)=L_0\frac{1}{p}x_0^p$ is the value function from (\ref
{eqPrimalProblemFinite}).
The following is contained in \cite{Nutz09a}, Lemma 3.5.
\begin{Lemma}\label{leBoundsForL} $L$ is a special semimartingale for
all $p$. If $p\in(0,1)$, then~$L,\allowbreak L_->0$, up to evanescence.
If $p<0$, the same holds provided that an optimal strategy exists.
\end{Lemma}
\begin{Prop}[(\cite{Nutz09a}, Proposition 3.4)]\label{prmartOptPrincipleForL}
Let $(\pi,c)\in\cA^{fE}$.
Then the process
\[
L_t\frac{1}{p}(X_t(\pi,c))^p + \int_0^t U_s(c_s) \mu
(ds), \qquad t\in[0,T],
\]
is a supermartingale; it is a martingale if and only if $(\pi,c)$ is optimal.
\end{Prop}

This is the ``martingale optimality principle.'' The expected terminal
value of this process
equals $E[\int_0^T U_t(c_t) \mu^\circ(dt)]$, hence,\vspace*{1pt} the assertion
fails for $(\pi,c)\in\cA\setminus\cA^{fE}$.

%

\subsection{Semimartingale characteristics}

In the remainder of this section we introduce tools which are necessary
to describe the optimization problem locally.
The use of semimartingale characteristics and set-valued processes
follows \cite{GollKallsen03} and~\cite{KaratzasKardaras07},\vadjust{\goodbreak}
which consider logarithmic utility and convex constraints. That problem
differs from ours in that it is ``myopic,'' that is,
the characteristics of $R$ are sufficient to describe the local problem and
so there is no need for an opportunity process.

We refer to \cite{JacodShiryaev03} for background regarding
semimartingale characteristics and random measures.
Let $\mu^R$ be the integer-valued random measure associated with the
jumps of $R$
and let $h\dvtx\R^d\to\R^d$ be a cut-off function, that is,
$h$ is bounded and $h(x)=x$ in a neighborhood of $x=0$.
Let $(B^R,C^R,\nu^R)$ be the predictable characteristics of $R$
relative to $h$.
The \textit{canonical representation} of $R$ (cf. \cite{JacodShiryaev03},
II.2.35) is
%
%
\begin{equation}\label{eqcanonicalRepR}
R=B^R + R^c + h(x)\ast(\mu^R-\nu^R) + \bigl(x-h(x)\bigr)\ast\mu^R.
\end{equation}
The finite variation process $(x-h(x))\ast\mu^R$ contains essentially
the ``large'' jumps of $R$. The rest is the canonical decomposition of
the special semimartingale $\bar{R}=R-(x-h(x))\ast\mu^R$, which has bounded jumps:
$B^R=B^R(h)$ is predictable of finite variation, $R^c$ is a continuous
local martingale and $h(x)\ast(\mu^R-\nu^R)$ is a purely discontinuous
local martingale.

As $L$ is a special semimartingale (Lemma \ref{leBoundsForL}), it has a
canonical decomposition $L=L_0+ A^L + M^L$. Here $L_0$ is constant,
$A^L$ is predictable of finite variation and also called the \textit
{drift} of $L$, $M^L$ is a local martingale and $A_0^L=M^L_0=0$.
Analogous notation will be used for other special semimartingales.
It is then possible to consider the characteristics $(A^L,C^L,\nu^L)$
of $L$ with respect to the identity instead of a cut-off function.
Writing $x'$ for the identity on $\R$, the canonical representation is
\[
L=L_0+ A^L + L^c + x' \ast(\mu^L-\nu^L)
\]
(see \cite{JacodShiryaev03}, II.2.38).
It will be convenient to use the joint characteristics of the $\R
^d\times\R$-valued process $(R,L)$.
We denote a generic point in $\R^d\times\R$ by $(x,x')$ and let
$(B^{R,L},C^{R,L},\nu^{R,L})$ be the
characteristics of $(R,L)$ with respect to the function $(x,x')\mapsto
(h(x),x')$.
More precisely, we choose ``good'' versions of the characteristics
so that they satisfy the properties given in \cite{JacodShiryaev03}, II.2.9.
For the $(d+1)$-dimensional process $(R,L)$ we have the canonical representation
\begin{eqnarray*}
\pmatrix{R \cr L }&=&
\pmatrix{0 \cr L_0 }+
\pmatrix{B^R \cr A^L }+
\pmatrix{R^c \cr L^c }+
\pmatrix{h(x) \cr x' } \ast(\mu^{R,L}-\nu^{R,L}) \\
&&{}+
\pmatrix{x-h(x) \cr0 } \ast\mu^{R,L}.
\end{eqnarray*}

We denote by $(b^{R,L},c^{R,L},F^{R,L};A)$ the \textit{differential}
characteristics with respect to
a predictable locally integrable increasing process $A$, for example,
\[
A_t:= t + \sum_{i} \Var(B^{RL,i})_t + \sum_{i,j} \Var(C^{RL,ij})_t+
\bigl(|(x,x')|^2\wedge1\bigr)\ast\nu_t^{R,L}.
\]
Then $b^{R,L}\sint A=B^{R,L}$, $c^{R,L}\sint A=C^{R,L}$ and
$F^{R,L}\sint A=\nu^{R,L}$. We shall write $b^{R,L}=(b^R,a^L)^\top$
and
$c^{R,L}=\left({c^R \enskip c^{RL}} \atop
{(c^{RL})^\top\enskip c^L} \right),
$ %
that is, $c^{RL}$ is a $d$-vector\vadjust{\goodbreak} satisfying $(c^{RL})\sint A = \langle
R^c,L^c \rangle.$
We will often use that
%
%
\begin{equation}\label{eqFRLintegrates}
\int_{\R^d\times\R} (|x|^2+|x'|^2)\wedge(1+|x'|)
F^{R,L}(d(x,x'))<\infty,
\end{equation}
because $L$ is a special semimartingale; cf. \cite{JacodShiryaev03}, II.2.29.
Let $Y$ be any scalar semimartingale with differential characteristics
$(b^Y,c^Y,F^Y)$ relative to $A$ and a cut-off function $\bar{h}$.
We call
\[
a^Y:=b^Y+\int\bigl(x-\bar{h}(x)\bigr) F^Y(dx)
\]
the \textit{drift rate}\vspace*{1pt} of $Y$ whenever the integral is well defined with
values in $[-\infty,\infty]$, even if it is not finite. Note that $a^Y$
does not depend on the choice of $\bar{h}$. If $Y$ is special, the
drift rate is finite and even $A$-integrable (and vice versa). As an
example, $a^L$ is the drift rate of $L$ and $a^L\sint A=A^L$ yields the drift.
\begin{Remark}\label{rkDriftRateNonnegSM}
Assume $Y$ is a \textup{nonpositive} scalar semimartingale. Then its
drift rate $a^Y$ is well defined with values in $[-\infty,\infty)$.
Indeed, the fact that
$Y=Y_-+\Delta Y\leq0$ implies that $x\leq-Y_-$, $F^Y(dx)$-a.e.
\end{Remark}

If $Y$ is a scalar semimartingale with drift rate $a^Y\in[-\infty,0]$,
we call $Y$ a~\textit{semimartingale with nonpositive drift rate}. Here
$a^Y$ need not be finite, as in the case of a compound Poisson process
with negative, nonintegrable jumps.
We refer to Kallsen \cite{Kallsen04} for the concept of $\sigma
$-localization. Denoting by $L(A)$ the set of $A$-integrable processes
and recalling that $\cF_0$ is trivial, we conclude the following, for example,
from \cite{KaratzasKardaras07}, Appendix 3.
\begin{Lemma}\label{lenonnegDriftRateEquivalences}
Let $Y$ be a semimartingale with nonpositive drift rate.
\begin{longlist}
\item$Y$ is a $\sigma$-supermartingale $\Leftrightarrow$ $a^Y$ is finite
$\Leftrightarrow$ $Y$ is $\sigma$-locally of class~(D).
\item$Y$ is a local supermartingale $\Leftrightarrow$ $a^Y\in L(A)$
$\Leftrightarrow$ $Y$ is locally of class~(D).
\item If $Y$ is uniformly bounded from below, it is a supermartingale.
\end{longlist}
\end{Lemma}

\subsection{Constraints and degeneracies}\label{seConstraintsDegeneracies}

We introduce some set-valued processes that will be used in the sequel,
that is, for each $(\omega,t)$
they describe a subset of~$\R^d$.
We refer to Rockafellar \cite{Rockafellar76} and Aliprantis and
Border \cite{AliprantisBorder06}, Section 18, for background.

We start by expressing the budget constraint in this fashion.
The process
\[
\sC^0_t(\omega):=\bigl\{y\in\R^d\dvtx F^R_t(\omega)\{x\in\R^d\dvtx
y^\top x < -1\}=0\bigr\}
\]
was called the \textit{natural constraints} in \cite{KaratzasKardaras07}.
Clearly $\sC^0$ is closed, convex and contains the origin. Moreover,
one can check (see \cite{KaratzasKardaras07}, Section 3.3) that it is
\textit{predictable} in the sense that for each closed $G\subseteq\R^d$,
the lower inverse image
$
(\sC^{0})^{-1}(G)=\{(\omega,t)\dvtx\sC_t(\omega)\cap G\neq
\varnothing\}
$
is predictable. (Here one can replace closed by compact or by open;
see \cite{Rockafellar76}, 1A.)
A statement such as ``$\sC^0$ is closed'' means that $\sC^0_t(\omega)$
is closed for all $(\omega,t)$; moreover, we will often omit the
arguments $(\omega,t)$.
We also consider the slightly smaller set-valued process
\[
\sC^{0,*}:=\bigl\{y\in\R^d\dvtx F^R\{x\in\R^d\dvtx y^\top x \leq-1\}
=0\bigr\}.
\]

These processes relate to the budget constraint as follows.
\begin{Lemma}
A process $\pi\in L(R)$ satisfies $\cE(\pi\sint R) \geq0$ $(>0)$ up
to evanescence if and only if $\pi\in\sC^0 (\sC^{0,*})$
$P\otimes A$-a.e.
\end{Lemma}
\begin{pf}
Recall that $\cE(\pi\sint R) > 0$ if and only if $1+\pi^\top\Delta
R>0$ (\cite{JacodShiryaev03},~II.8a). Writing
$V(x)=1_{\{x\dvtx1+\pi^\top x\leq0\}}(x)$, we have $(P\otimes
A)\{\pi\notin\sC^{0,*}\}=\break E[V(x)\ast\nu^R_T]=E[V(x)\ast
\mu^R_T]=E[\sum_{s\leq T} 1_{\{x\dvtx1+\pi_s ^\top\Delta R_s\leq
0\}}]$. For the\vspace*{1pt} equivalence with~$\sC^0$, interchange strict and
nonstrict inequality signs.
\end{pf}

The process $\sC^{0,*}$ is not closed in general (nor relatively open).
Clearly, we have $\sC^{0,*}\subseteq\sC^0$, and in fact $\sC^0$ is the
closure of
$\sC^{0,*}$; for $y\in\sC_t^0(\omega)$, the sequence $\{(1-1/n)y\}
_{n\geq1}$ is in $\sC_t^{0,*}(\omega)$ and converges to~$y$.
This implies that~$\sC^{0,*}$ is predictable; cf. \cite
{AliprantisBorder06}, 18.3.
We will not be able to work directly with~$\sC^{0,*}$ because
closedness is essential for the measurable selection arguments that
will be used.

We turn to the exogenous portfolio constraints, that is, the
set-valued process $\sC$ containing the origin. We consider the
following conditions:
\begin{longlist}[(C3)]
\item[(C1)] $\sC$ is predictable.
\item[(C2)] $\sC$ is closed.
\item[(C3)] \textit{If} $p\in(0,1)$: there exists a $(0,1)$-valued
process $\underline{\eta}$ such that
\[
y\in(\sC\cap\sC^{0}) \setminus\sC^{0,*} \Longrightarrow\eta y \in\sC
\qquad\mbox{for all } \eta\in(\underline{\eta},1), P\otimes
A\mbox{-a.e.}
\]
\end{longlist}
Condition (C3) is clearly satisfied if $\sC\cap\sC^{0} \subseteq\sC
^{0,*}$, which includes the case
of a continuous process $R$, and it is always satisfied if $\sC$ is
convex or, more generally, star-shaped with respect to the origin. If
$p<0$, (C3) should be read as always being satisfied.

We require (C3) to exclude a degenerate situation where, despite the
Inada condition $U'(0)=\infty$, it is actually desirable for the agent
to have a wealth process that vanishes in some states. That situation,
illustrated in the subsequent example, would necessitate a more
complicated notation while it can arise only in cases that are of minor
interest.
\begin{Example}
We assume that there is no intermediate consumption and $x_0=1$.
Consider the one-period binomial model of a financial market, that is,
$S=\cE(R)$ is a scalar process which is constant up to time $T$, where
it has a single jump,\vadjust{\goodbreak} say $P[\Delta R_T=-1]=p_0$ and $P[\Delta
R_T=K]=1-p_0$, where $K>0$ is a constant and $p_0\in(0,1)$. The
filtration is generated by~$R$ and we consider $\sC\equiv\{0\}\cup\{
1\}$.
Then $E[U(X_T(\pi))]=U(1)$ if $\pi_T=0$ and
$E[U(X_T(\pi))]=p_0U(0)+(1-p_0)U(1+K)$
if $\pi_T=1$.
If $U(0)>-\infty$, and if~$K$ is large enough, $\pi_T=1$ performs
better despite the fact that
its \textit{terminal wealth vanishes} with probability $p_0>0$. Of
course, this cannot happen if $U(0)=-\infty$, that is, $p<0$.
\end{Example}

By adjusting the constants in the example, one can also see that under
nonconvex constraints, there is in general \textit{no uniqueness} for the
optimal wealth processes (even if they are positive).

The final set-valued process is related to linear dependencies of the assets.
As in \cite{KaratzasKardaras07}, the predictable process of
\textit{null-investments} is
\[
\sN:=\bigl\{y\in\R^d\dvtx y^\top b^R=0, y^\top c^R=0, F^R\{x\dvtx y^\top
x\neq0\}=0 \bigr\}.
\]
Its values are linear subspaces of $\R^d$, hence closed, and provide
the pointwise description of the null-space of $H\mapsto H\sint R$.
That is,
$H\in L(R)$ satisfies $H\sint R\equiv0$ if and only if $H\in\sN$
$P\otimes A$-a.e.
An investment with values in $\sN$ has no effect on the wealth process.
%

\section{The Bellman equation}\label{sebellmanEqn}

We have now introduced the necessary notation to formulate our first
main result.
Two special cases of our Bellman equation can be found in the
pioneering work of Mania and Tevzadze \cite{ManiaTevzadze03}
and Hu, Imkeller and M\"uller \cite{HuImkellerMuller05}. These
articles consider models with continuous asset prices and we shall
indicate the connections as we specialize to that case in Section \ref
{secontCase}.
A~related equation also arises in the study of mean--variance hedging by
\v{C}ern\'y and Kallsen \cite{CernyKallsen07} in the
context of locally square-integrable semimartingales, although they do
not use dynamic programming explicitly. Due to the quadratic setting,
that equation is more explicit than ours and the mathematical treatment
is quite different.
Czichowsky and Schweizer \cite{CzichowskySchweizer09b} study a
cone-constrained version of the related Markowitz problem and there the
equation is no longer explicit.

The Bellman equation highlights the local structure of our utility
maximization problem. In addition, it has two main benefits.
First, it can be used as an abstract tool to derive properties of the
optimal strategies and the opportunity process
(e.g., Nutz \cite{Nutz09d}).
Second, one can try to solve the equation directly in a given model and
to deduce the optimal strategies. This is the point of view taken in
Section \ref{severification} and obviously requires the precise form of
the equation.

The following assumptions are in force for the entire Section
\ref{sebellmanEqn}.
\begin{Assumptions}\label{asbellmanSection}
The value of the utility maximization problem is finite, there exists
an optimal strategy $(\hpi,\hc)\in\cA$ and $\sC$ satisfies (C1)--(C3).
\end{Assumptions}

\subsection{Bellman equation in joint characteristics}

Our first main result is the Bellman equation stated as a
description of the drift rate of the opportunity process.
We recall the conjugate function $U_t^*(y)=-\frac{1}{q}y^{q}D_t^{\beta}$.
\begin{Thm}\label{thBellmanEqnCharact}
The drift rate $a^L$ of the opportunity process satisfies
%
%
\begin{equation}\label{eqLDriftRate}
-p^{-1} a^L = U^*(L_-) \,\frac{d\mu}{dA} + \max_{y\in\sC\cap\sC^0} g(y),
\end{equation}
where $g$ is the predictable random function
%
%
\begin{eqnarray}\label{eqDefOfg}
g(y)
& := & L_{-}y^\top\biggl( b^R + \frac{c^{RL}}{L_{-}}+ \frac{(p-1)}{2}
c^R y \biggr)\nonumber\\
&&{} + \int_{\R^d\times\R} x' y^\top h(x) F^{R,L}(d(x,x'))
\nonumber\\[-8pt]\\[-8pt]
&&{} + \int_{\R^d\times\R} (L_-+x') \{p^{-1}(1+y^\top x)^p\nonumber\\
&&\hspace*{88.46pt}{} - p^{-1} -
y^\top h(x)\} F^{R,L}(d(x,x')).\nonumber
\end{eqnarray}
The unique ($P\otimes\mu^\circ$-a.e.) optimal propensity to consume is
%
%
\begin{equation}\label{eqoptPropConsGen}
\hkappa=\biggl(\frac{D}{L}\biggr)^{{1}/({1-p})}.
\end{equation}
Any optimal trading strategy $\pi^*$ satisfies
%
%
\begin{equation}\label{eqoptStrategyGen}
\pi^*\in\mathop{\argmax}_{\sC\cap\sC^0} g,
\end{equation}
and the corresponding optimal wealth process and consumption are given by
\[
X^*=x_0\cE(\pi^*\sint R - \hkappa\sint\mu);\qquad
c^*=X^*\hkappa.
\]
\end{Thm}

We shall see in the proof that the maximization in (\ref{eqLDriftRate})
can be understood as a local version of the optimization problem.
Indeed, recalling (\ref{eqconvexConjugate}), the right-hand side
of (\ref{eqLDriftRate}) is the maximum of a single function over
certain points $(k,y)\in\R_+\times\R^d$ that correspond to the
admissible controls $(\kappa,\pi)$. Moreover, optimal controls are
related to maximizers of this function, a~characteristic feature of any
dynamic programming equation. The maximum of~$g$ is not explicit due to
the jumps of~$R$; this simplifies
in the continuous case considered in Section \ref{secontCase} below.
Some mathematical comments are also in order.
\begin{Remark}\label{rkAfterBellmanThm}
\textup{(i)} The random function $g$ is well defined on $\sC^0$ in the
extended sense (see Lemma \ref{lePropertiesOfI})
and it does not depend on the choice of the cut-off function $h$
by \cite{JacodShiryaev03}, \textup{II.2.25.}

\textup{(ii)} For $p<0$ we have a more precise statement: given $\pi^*\in L(R)$
and~$\hkappa$ as in~(\ref{eqoptPropConsGen}), $(\pi^*,\hkappa)$ is optimal
\textup{if and only if} $\pi^*$ takes values in $\sC\cap\sC^0$ and
maximizes~$g$. This will follow from Corollary \ref
{codirectVerificationBoundedL} applied to the triplet~$(L,\pi
^*,\hkappa)$.

\textup{(iii)} For $p\in(0,1)$, partial results in this direction follow from
Section \ref{severification}.
The question is trivial for convex $\sC$ by the next item.

\textup{(iv)} If $\sC$ is convex, $\argmax_{\sC\cap\sC^0}g$ is unique in the
sense that the difference of any two elements lies in $\sN$ (see
Lemma \ref{legConcave}).
\end{Remark}

We split the proof of Theorem \ref{thBellmanEqnCharact} into several
steps; the plan is as follows. Let $(\pi,\kappa)\in\cA^{fE}$ and denote
$X=X(\pi,\kappa)$.
We recall from Proposition \ref{prmartOptPrincipleForL} that
\[
Z(\pi,\kappa):=L\frac{1}{p}X^p + \int U_s(\kappa_s X_s) \mu(ds)
\]
is a supermartingale, and a martingale if and only if $(\pi,\kappa)$ is
optimal. Hence, we shall calculate its drift rate and then maximize
over $(\pi,\kappa)$; the maximum will be attained at any optimal
strategy. This is fairly straightforward and essentially the content of
Lemma \ref{lehpiOverPredProc} below.
In the Bellman equation, we maximize over a subset of $\R^d$ for each
$(\omega,t)$ and not over a set of strategies.
This final step is a measurable selection problem and its solution will
be the second part of the proof.
\begin{Lemma}\label{leCompensatorJointChar}
Let $(\pi,\kappa)\in\cA^f$. The drift rate of $Z(\pi,\kappa)$ is
\[
a^{Z(\pi,\kappa)}=X(\pi,\kappa)^{p}_- \biggl( p^{-1} a^L + f(\kappa
)\,\frac
{d\mu}{dA} + g(\pi)\biggr) \in[-\infty,\infty),
\]
where\vspace*{2pt} $f_t(k):= U_t(k) - L_{t-}k$ and $g$ is given by (\ref{eqDefOfg}).
Moreover, $a^{Z(\hpi,\hkappa)}=0$ and $a^{Z(\pi,\kappa)}\in
(-\infty
,0]$ for $(\pi,\kappa)\in\cA^{fE}$.
\end{Lemma}
\begin{pf}
We can assume that the initial capital is $x_0=1$. Let $(\pi,\kappa
)\in
\cA^f$, then in particular $Z:=Z(\pi,\kappa)$ is finite. We also set
$X:=X(\pi,\kappa)$. By It\^o's formula, we have $X^p=\cE(\pi\sint R -
\kappa\sint\mu)^p=\cE(Y)$ with
\begin{eqnarray*}
Y&=&p(\pi\sint R - \kappa\sint\mu) + \frac{p(p-1)}{2}\pi^\top
c^R\pi
\sint A\\
&&{} + \{(1+\pi^\top x)^p-1-p\pi^\top x\}\ast\mu^R.
\end{eqnarray*}
Integrating by parts in the definition of $Z$ and using $X_s=X_{s-}$
$\mu(ds)$-a.e. (path-by-path), we have
$X_-^{-p}\sint Z = p^{-1}(L-L_0 +L_-\sint Y + [L,Y])+ U(\kappa)\sint
\mu$. Here
\begin{eqnarray*}
[L,Y]&=&[L^c,Y^c]+ \sum\Delta L \Delta Y \\
&=& p\pi^\top c^{RL}\sint A + px'\pi^\top x \ast\mu^{R,L}\\
&&{} + x'\{(1+\pi^\top x)^{p} - 1 - p\pi^\top x\} \ast\mu^{R,L}.
\end{eqnarray*}
Thus $X_-^{-p}\sint Z$ equals
\begin{eqnarray*}
&& p^{-1}(L-L_0) + L_-\pi\sint R + f(\kappa)\sint\mu\\
&&\qquad{} + L_- \frac
{(p-1)}{2} \pi^\top c^R \pi\sint A + \pi^\top c^{RL}\sint A
+ x' \pi^\top x \ast\mu^{R,L}\\
&&\qquad{} + (L_-+x')\{p^{-1}(1+\pi^\top x)^p
- p^{-1} - \pi^\top x\} \ast\mu^{R,L}.
\end{eqnarray*}
Writing $x=h(x)+x-h(x)$ and $\bar{R}=R- (x-h(x))\ast\mu^R$ as in
(\ref{eqcanonicalRepR}),
%
%
\begin{eqnarray}\label{eqproofCompensatorJointChar}
X_-^{-p}\sint Z &=&
p^{-1}(L-L_0) + L_-\pi\sint\bar{R} + f(\kappa)\sint\mu\nonumber\\
&&{}+ L_- \pi
^\top\biggl(\frac{c^{RL}}{L_-} + \frac{(p-1)}{2}c^R \pi\biggr)\sint
A + x'\pi^\top h(x) \ast\mu^{R,L}\\
&&{} + (L_-+x')\{p^{-1}(1+\pi^\top
x)^p - p^{-1} - \pi^\top h(x)\} \ast\mu^{R,L}.\nonumber
\end{eqnarray}
Since $\pi$ need not be locally bounded, we use from now on a
predictable cut-off function $h$ such that $\pi^\top h(x)$ is bounded,
for example, $h(x)=x1_{\{|x|\leq1\}\cap\{|\pi^\top x|\leq1\}}$.
Then the compensator of $x'\pi^\top h(x) \ast\mu^{R,L}$ exists, since
$L$ is special. %

Let $(\pi,\kappa)\in\cA^{fE}$. Then
the compensator of the last integral in the right-hand side of (\ref
{eqproofCompensatorJointChar}) also exists; indeed, all other terms in
that equality are special, since $Z$ is a supermartingale. The drift
rate can now be read from~(\ref{eqproofCompensatorJointChar}) and
(\ref{eqcanonicalRepR}), and it is nonpositive by the supermartingale
property. The drift rate vanishes for the optimal $(\hpi,\hkappa)$ by
the martingale condition from Proposition \ref{prmartOptPrincipleForL}.

Now consider $(\pi,\kappa)\in\cA^f\setminus\cA^{fE}$. Note that
necessarily $p<0$ (otherwise $\cA^f=\cA^{fE}$). Thus $Z\leq0$, so by
Remark \ref{rkDriftRateNonnegSM} the drift rate $a^Z$ is well defined
with values in $[-\infty,\infty)$---alternatively, this can also be
read from the integrals in (\ref{eqproofCompensatorJointChar})
via~(\ref{eqFRLintegrates}). Using directly the definition of $a^Z$, we
find the same formula for $a^Z$ is as above.
\end{pf}

We do not have the supermartingale property for $(\pi,\kappa)\in\cA
^f\setminus\cA^{fE}$, so it is not evident that $a^{Z(\pi,\kappa
)}\leq
0$ in that case. However, we have the following.

\begin{Lemma}\label{leZdriftrateNotPositive}
Let $(\pi,\kappa)\in\cA^f$. Then $a^Z(\pi,\kappa)\in[0,\infty]$
implies \mbox{$a^Z(\pi,\kappa)=0$}.
\end{Lemma}

\begin{pf}
Denote $Z=Z(\pi,\kappa)$. For $p>0$ we have $\cA^f=\cA^{fE}$ and the
claim is immediate from Lemma \ref{leCompensatorJointChar}.
Let $p<0$. Then $Z\leq0$ and in view of Lemma~\ref
{lenonnegDriftRateEquivalences}(iii), $a^Z\in[0,\infty]$ implies that
$Z$ is a submartingale. Therefore, we have that $E[Z_T]=E[\int_0^T U_t(\kappa
_tX_t(\pi,\kappa)) \mu^\circ(dt)]>-\infty$, that is, $(\pi,\kappa
)\in\cA^{fE}$.
Now Lemma~\ref{leCompensatorJointChar} yields $a^Z(\pi,\kappa)\leq0$.
\end{pf}

We observe in Lemma \ref{leCompensatorJointChar} that the drift rate
splits into separate functions involving $\kappa$ and $\pi$,
respectively. For this reason, we can single out the following proof:
\begin{pf*}{Proof of the consumption formula (\ref{eqoptPropConsGen})}
Let $(\pi,\kappa)\in\cA$. Note the following feature of our parametrization:
we have $(\pi,\kappa^*)\in\cA$ for \textit{any}
nonnegative optional process $\kappa^*$ such that $\int_0^T \kappa
_s^* \mu(ds)<\infty$ and $\kappa^*_T=1$.
Indeed, the process $X(\pi,\kappa)=x_0\cE(\pi\sint R - \kappa\sint\mu)$ is
positive by assumption. As~$\mu$ is continuous,
$X(\pi,\kappa^*)=x_0\cE(\pi\sint R - \kappa^*\sint\mu)$ is also positive.

In particular, let $(\hpi,\hkappa)$ be optimal, $\beta=(1-p)^{-1}$ and
$\kappa^*=(D/L)^\beta$, then
$(\hpi,\kappa^*)\in\cA$. In fact, the paths of
$U(\kappa^*X(\hpi,\kappa^*))=p^{-1}D^{\beta p+1}X(\hpi,\kappa
^*)^pL^{-\beta p}$ are bounded $P$-a.s.
(because the processes are c\`adl\`ag; $L,L_->0$, and $\beta p+1=\beta
>0$) so that
$(\hpi,\kappa^*)\in\cA^f$.

Note that $P\otimes\mu$-a.e., we have $\kappa^*=(D/L_-)^\beta
=\argmax
_{k\geq0} f(k)$, hence, $f(\kappa^*)\geq f(\hkappa)$. Suppose\vspace*{2pt}
$(P\otimes\mu)\{f(\kappa^*)>f(\hkappa)\}>0$, then
the formula from Lemma \ref{leCompensatorJointChar} and $a^{Z(\hpi
,\hkappa)}=0$ imply
$a^{Z(\hpi,\kappa^*)}\geq0$ and $(P\otimes A)\{a^{Z(\hpi,\kappa
^*)}>0\}
>0$, a~contradiction to Lemma \ref{leZdriftrateNotPositive}.
It follows that $\hkappa=\kappa^*$ $P\otimes\mu$-a.e. since $f$ has a
unique maximum.
\end{pf*}

\begin{Remark}
$\!\!\!$The previous proof does not use the assumptions \mbox{\textup{(C1)--(C3)}}.
\end{Remark}
\begin{Lemma}\label{lehpiOverPredProc}
Let $\pi$ be a predictable process with values in $\sC\cap\sC
^{0,*}$. Then
\[
(P\otimes A)\{g(\hpi)<g(\pi)\}=0.
\]
\end{Lemma}
\begin{pf}
We argue by contradiction and assume $(P\otimes A) \{g(\hpi)<\break g(\pi)\}
>0$. By redefining $\pi$, we may assume that $\pi=\hpi$ on the
complement of this predictable set.
Then
%
%
\begin{equation}\label{eqProofPosMeasure}
g(\hpi)\leq g(\pi) \quad\mbox{and}\quad (P\otimes A) \{g(\hpi)<g(\pi)\}>0.
\end{equation}
Using that~$\pi$ is $\sigma$-bounded, %
we can find a constant $C>0$ such that the process
$\tilde{\pi}:=\pi1_{|\pi|\leq C} + \hpi1_{|\pi|> C}$ again
satisfies (\ref{eqProofPosMeasure}), that is,
we may assume that~$\pi$ is $R$-integrable. Since $\pi\in\sC\cap
\sC
^{0,*}$, this implies $(\pi,\hkappa)\in\cA$
(as observed above, the consumption $\hkappa$ plays no role here). The
contradiction follows as in the previous proof.
\end{pf}

In view of Lemma \ref{lehpiOverPredProc}, the main task will be to
construct a \textit{measurable} maximizing sequence for $g$.
\begin{Lemma}\label{lemeasmaxsequence}
Under Assumptions \ref{asbellmanSection}, there exists a sequence
$(\pi
^{n})$ of predictable $\sC\cap\sC^{0,*}$-valued
processes such that
\[
\limsup_n g(\pi^n) = \sup_{\sC\cap\sC^0} g,\qquad P\otimes A\mbox{-a.e.}
\]
\end{Lemma}

We defer the proof of this lemma to Appendix \ref{seMeasSelection},
together with the study of the properties of $g$. The theorem can then
be proved as follows.

\begin{pf*}{Proof of Theorem \ref{thBellmanEqnCharact}}
Let $\pi^n$ be as in Lemma \ref{lemeasmaxsequence}. Then Lemma~\ref
{lehpiOverPredProc}, with $\pi=\pi^n$, yields
$g(\hpi)= \sup_{\sC\cap\sC^0} g$, which is (\ref{eqoptStrategyGen}).
By Lemma \ref{leCompensatorJointChar} we have
$0=a^{Z(\hpi,\hkappa)}=p^{-1} a^L + f(\hkappa)\,\frac{d\mu}{dA} +
g(\hpi
)$. This is (\ref{eqLDriftRate}) as
$f(\hkappa)=U^*(L_-)$ holds $P\otimes\mu$-a.e. due to (\ref{eqoptPropConsGen}).
\end{pf*}

\subsection{Bellman equation as BSDE}

In this section we express the Bellman equation as a BSDE.
The unique orthogonal decomposition of the local martingale $M^L$ with
respect to $R$; cf. \cite{JacodShiryaev03}, III.4.24
leads to the representation
%
%
\begin{equation}\label{eqGeneralGKWdecompL}
L= L_0 + A^L + \varphi^L \sint R^c + W^L\ast(\mu^R-\nu^R) + N^L,
\end{equation}
where, using the notation of \cite{JacodShiryaev03}, $\varphi^L\in
L^2_{\mathrm{loc}}(R^c)$, $W^L\in G_{\mathrm{loc}}(\mu^R)$, and $N^L$ is a~local
martingale such that $\langle(N^L)^c, R^c \rangle=0$ and $M^P_{\mu
^R}(\Delta N^L|
\widetilde{\cP})=0$. The last statement means that $E[(V \Delta
N^L)\ast\mu^R_T]=0$ for any sufficiently integrable predictable
function $V=V(\omega,t,x)$.
We also introduce
\[
\widehat{W}^L_t :=\int_{\R^d} W^L(t,x) \nu^R(\{t\}\times dx),
\]
then\vspace*{1pt} $\Delta(W^L\ast(\mu^R-\nu^R)) = W^L(\Delta R) 1_{\{
\Delta R\neq0\}}-\widehat{W}^L$
by definition of the purely discontinuous local martingale $W^L\ast
(\mu
^R-\nu^R)$ and we can write
\[
\Delta L = \Delta A^L + W^L(\Delta R) 1_{\{\Delta R\neq0\}}-\widehat
{W}^L + \Delta N^L.
\]
We recall that Assumptions \ref{asbellmanSection} are in force.
Now (\ref{eqLDriftRate}) can be restated as follows, the random
function $g$ being the same as before but in new notation.

\begin{Cor}\label{coBellmanBsdeGen}
The opportunity process $L$ and the processes defined by (\ref
{eqGeneralGKWdecompL}) satisfy the BSDE
%
%
\begin{eqnarray}\label{eqBellmanBsdeGenForL}
L &=& L_0 -p U^*(L_-)\sint\mu- p \max_{y\in\sC\cap\sC^0} g(y)\sint
A \nonumber\\[-8pt]\\[-8pt]
&&{}+
\varphi^L \sint R^c + W^L\ast(\mu^R-\nu^R) + N^L\nonumber
\end{eqnarray}
with terminal condition $L_T=D_T$,
where $g$ is given by
\begin{eqnarray*}%
g(y) &:=&
L_{-}y^\top\biggl( b^R + c^R \biggl(\frac{\varphi^L}{L_{-}} + \frac
{(p-1)}{2}y \biggr) \biggr)\\
&&{} + \int_{\R^d} \bigl(\Delta A^L+ W^L(x)-\widehat
{W}^L\bigr) y^\top h(x) F^R(dx) \\
&&{} + \int_{\R^d} \bigl(L_- + \Delta A^L+ W^L(x)-\widehat
{W}^L\bigr) \{p^{-1}(1+y^\top x)^p - p^{-1} - y^\top h(x)\}\\
&&\hspace*{27pt}{}\times F^R(dx).
\end{eqnarray*}
\end{Cor}

We observe that the orthogonal part $N^L$ does not appear in the
definition of~$g$. In a suitable setting, it is linked to the
``dual problem'' (see Re\-mark~\ref{rklossOfMass}).

It is possible (but notationally more cumbersome) to prove a version of
Lem\-ma~\ref{leCompensatorJointChar} using $g$ as in Corollary \ref
{coBellmanBsdeGen} and the decomposition (\ref{eqGeneralGKWdecompL}),
thus involving only the characteristics of $R$ instead of the joint
characteristics of $(R,L)$. Using this approach, we see that the
increasing process $A$ in the BSDE can be chosen based on $R$ and
without reference to $L$. This is desirable if we want to consider
other solutions of the equation, as in Section \ref{seminimality}.
One consequence is that $A$ can be chosen to be continuous if and only
if $R$ is quasi left-continuous; cf. \cite{JacodShiryaev03}, II.2.9.
Since $p^{-1} A^L =- f(\hkappa)\sint\mu- g(\hpi)\sint A$, $\Var(A^L)$
is absolutely continuous with respect to $A+\mu$, and we conclude the
following.
\begin{Remark}\label{rkALcontForSqlc}
If $R$ is quasi left-continuous, $A^L$ is continuous.
\end{Remark}

If $R$ is quasi left-continuous, $\nu^R(\{t\}\times\R^d)=0$ for all
$t$ by \cite{JacodShiryaev03}, II.1.19;
hence, $\widehat{W}^L=0$ and we have the simpler formula
\begin{eqnarray*}
g(y) &=& L_{-}y^\top\biggl( b^R + c^R \biggl(\frac{\varphi^L}{L_{-}} +
\frac{(p-1)}{2}y \biggr) \biggr) + \int_{\R^d} W^L(x) y^\top h(x) F^R(dx)
\\
&&{} + \int_{\R^d} \bigl(L_- + W^L(x)\bigr) \{p^{-1}(1+y^\top x)^p -
p^{-1} - y^\top h(x)\} F^R(dx).
\end{eqnarray*}

\subsection{The case of continuous prices}\label{secontCase}

In this section we specialize the previous results to the case where
$R$ is a continuous semimartingale and mild additional conditions are
satisfied. %
As usual in this setting, the martingale part of $R$ will be denoted by
$M$ rather than $R^c$. In addition to Assumptions~\ref
{asbellmanSection}, the following conditions are in force for the
present Section~\ref{secontCase}.

\begin{Assumptions}\label{ascontCase}
\begin{longlist}
\item $R$ is continuous,
\item $R=M +\int d\langle M \rangle\lambda$ for some $\lambda\in
L^2_{\mathrm{loc}}(M)$
(\textit{structure condition}),
\item the orthogonal projection of $\sC$ onto $\sN^\bot$ is closed.
\end{longlist}
\end{Assumptions}

Note that $\sC^{0,*}=\R^d$ due to (i), in particular (C3) is void. When
$R$ is continuous, it necessarily satisfies (ii) when a
no-arbitrage property holds (see Schweizer \cite{Schweizer95b}).
By (i) and (ii) we can write the differential characteristics of~$R$
with respect to, for example, $A_t:=t+\sum_{i=1}^d \langle M^i \rangle_t$.
It will be convenient to factorize $c^R=\sigma\sigma^\top$, where
$\sigma$ is a predictable matrix-valued process, hence, $\sigma\sigma
^\top dA=d\langle M \rangle$.
Then (ii)\vspace*{1pt} implies $\sN=\ker\sigma^\top$ because $\sigma\sigma
^\top
y=0$ implies $(\sigma^\top y)^\top(\sigma^\top y)=0$. Since $\sigma
^\top
\dvtx\ker(\sigma^\top)^\bot\to\sigma^\top\R^d$ is a\vadjust{\goodbreak}
homeomorphism, we
see that (iii) is equivalent to
\[
\sigma^\top\sC\mbox{ is closed.}
\]
This condition depends on the semimartingale $R$. It is equivalent to
the closedness of $\sC$ itself if $\sigma$ has full rank.
For certain constraint sets (e.g., closed polyhedral or compact), the
condition is satisfied for \textit{all} matrices $\sigma$, but not so,
for example, for nonpolyhedral cone constraints. We mention that
violation of (iii) leads to nonexistence of optimal strategies in
simple examples; cf. \cite{Nutz09c}, Example 3.5, and we refer to
Czichowsky and Schweizer~\cite{CzichowskySchweizer09a}
for background. %

Under (i), (\ref{eqGeneralGKWdecompL}) is the more usual
Kunita--Watanabe decomposition
\[
L=L_0 + A^L + \varphi^L\sint M + N^L,
\]
where $\varphi^L\in L^2_{\mathrm{loc}}(M)$ and $N^L$ is a local martingale such
that $[M,N^L]=0$ (see Ansel and Stricker \cite{AnselStricker03},
Case 3).
If $\varnothing\neq K\subseteq\R^d$ is a closed set, we denote the
Euclidean distance to $K$ by $d_K(x)=\min\{|x-y|\dvtx y\in K\}$, and
$d^2_K$ is the squared distance. We also define the (set-valued) projection
$\Pi^{K}$ which maps $x\in\R^d$ to the points in $K$ with minimal
distance to $x$,
\[
\Pi^K(x)=\{y\in K\dvtx|x-y|=d_K(x)\}\neq\varnothing.
\]
If $K$ is convex, $\Pi^{K}$ is the usual (single-valued) Euclidean projection.
In the present continuous setting, the random function $g$
simplifies to
%
%
\begin{equation}\label{eqDefOfgContinuous}
g(y)=L_{-} y^\top\sigma\sigma^\top\biggl(\lambda+\frac{\varphi
^L}{L_{-}}+\frac{p-1}{2} y\biggr),
\end{equation}
and so the Bellman BSDE becomes more explicit.
\begin{Cor}\label{coBellmanBsdeCont}
Any optimal trading strategy $\pi^*$ satisfies
\[
\sigma^\top\pi^* \in\Pi^{\sigma^\top\sC} \biggl\{\sigma^\top
(1-p)^{-1}\biggl(\lambda+\frac{\varphi^L}{L_{-}}\biggr)\biggr\}.
\]
The opportunity process satisfies the BSDE
\[
L = L_0 -p U^*(L_-)\sint\mu+ F(L_-,\varphi^L)\sint A + \varphi
^L\sint
M + N^L;\qquad L_T=D_T,
\]
where
\begin{eqnarray*}
F(L_-,\varphi^L)&=&\frac{1}{2}L_{-} \biggl\{p(1-p) d^2_{\sigma^\top\sC
}\biggl(\sigma^\top(1-p)^{-1}\biggl(\lambda+\frac{\varphi^L}{L_{-}}
\biggr)\biggr)\\
&&\hspace*{99.4pt}{}
+\frac{p}{p-1} \biggl|\sigma^\top\biggl(\lambda+\frac{\varphi
^L}{L_{-}}\biggr)\biggr|^2 \biggr\}.
\end{eqnarray*}
If $\sC$ is a convex cone,
$
F(L_-,\varphi^L)=\frac{p}{2(p-1)}L_{-} |\Pi^{\sigma^\top\sC} \{
\sigma^\top(\lambda+\frac{\varphi^L}{L_{-}})\}|^2
$.
If $\sC=\R^d$, then
$
F(L_-,\varphi^L)\sint A=\frac{p}{2(p-1)}\int L_{-}(\lambda+\frac
{\varphi^L}{L_{-}})^\top \,d\langle M \rangle(\lambda+\frac{\varphi
^L}{L_{-}})
$
and the unique (mod. $\sN)$ optimal trading strategy is $\pi
^*=(1-p)^{-1}(\lambda+\frac{\varphi^L}{L_{-}})$.
\end{Cor}
\begin{pf}
Let $\beta=(1-p)^{-1}$. Then $\sigma^\top(\argmax_\sC g) = \Pi
^{\sigma^\top\sC} \{\sigma^\top\beta(\lambda+\frac{\varphi
^L}{L_{-}})\}$ by completing the square in (\ref{eqDefOfgContinuous}),
moreover, for any $\pi^*\in\argmax_\sC g$,
\[
g(\pi^*)=\frac{1}{2}L_{-} \biggl\{
\beta\biggl(\lambda+\frac{\varphi^L}{L_{-}}\biggr)^\top\sigma\sigma^\top
\biggl(\lambda+\frac{\varphi^L}{L_{-}}\biggr)
- \beta^{-1} d^2_{\sigma^\top\sC}\biggl(\sigma^\top\beta\biggl(\lambda
+\frac{\varphi^L}{L_{-}}\biggr)\biggr)
\biggr\}.
\]
In the case where $\sC$, and hence $\sigma^\top\sC$, is a convex
cone, $\Pi:=\Pi^{\sigma^\top\sC}$ is single-valued, positively
homogeneous, and
$\Pi x$ is orthogonal to $x - \Pi x$ for any~$x$ in $\R^d$. Writing
$\Psi:=\sigma^\top(\lambda+\frac{\varphi^L}{L_{-}})$ we get
$
g(\pi^*) =L_{-} \beta(\Pi\Psi)^\top(\Psi- \frac{1}{2} \Pi\Psi)
= L_{-} \frac{1}{2}\beta(\Pi\Psi)^\top(\Pi\Psi).
$
Finally, $\Pi\Psi=\Psi$ if $\sC=\R^d$. The result follows from
Corollary \ref{coBellmanBsdeGen}.
\end{pf}

Of course the consumption formula (\ref{eqoptPropConsGen}) and
Remark \ref{rkAfterBellmanThm} still apply.
We remark that the BSDE for the unconstrained case $\sC=\R^d$ (and
$\mu
=0$, $D=1$) was previously obtained in \cite{ManiaTevzadze03} in a
similar spirit.
A variant of the constrained BSDE for an It\^o process model (and $\mu
=0$, $D=1$) appears in \cite{HuImkellerMuller05},
where a converse approach is taken: the equation is derived only
formally and then existence results for BSDEs are employed together
with a verification argument. We shall extend that result in
Section \ref{severification} (Example \ref{exHuImkellerMuller}) when we
study verification.

If $L$ is continuous, the BSDE of Corollary \ref{coBellmanBsdeCont}
simplifies if it is stated for $\log(L)$ rather than $L$, but in
general the given form is more convenient as the jumps are ``hidden''
in $N^L$.
\begin{Remark}\label{rkLcontAndBSDEAppl}
\textup{(i)} Continuity of $R$ does not imply that $L$ is continuous. For
instance, in the It\^o process model of Barndorff-Nielsen and
Shephard~\cite{BarndorffNielsenShephard01}
with L\'evy driven coefficients, the opportunity process is not
continuous (see, e.g.,
Theorem~3.3 and the subsequent remark in Kallsen and Muhle-Karbe~\cite
{KallsenMuhleKarbe08}). If $R$ satisfies the structure condition and
the filtration $\F$ is continuous, it clearly follows that $L$ is continuous.
Here $\F$ is called continuous if all $\F$-martingales are continuous,
as, for example, for the Brownian filtration.
In general, $L$ is related to the predictable characteristics of the
asset returns rather than their levels.
As an example, L\'evy models have jumps but constant characteristics;
here $L$ turns out to be a smooth function (see~\cite{Nutz09c}).

\textup{(ii)} In the present setting we see that $F$ has quadratic growth in
$\varphi^L$, so that the Bellman equation is a ``quadratic BSDE''
(see also Example~\ref{exHuImkellerMuller}). In general, $F$ does not
satisfy the bounds which are
usually assumed in the theory of such BSDEs. Together with existence
results for the utility maximization problem
(see the citations from the \hyperref[intro]{Introduction}),
the Bellman equation yields various examples of BSDEs with the
opportunity process as a solution. This includes terminal conditions
$D_T$ which are integrable and unbounded (see also \cite{Nutz09a},
Remark 2.4).
\end{Remark}

\section{Minimality of the opportunity process}\label{seminimality}%

This section considers the Bellman equation as such, having possibly
many solutions, and we characterize the opportunity process
as the minimal solution.
As mentioned above, it seems more natural to use the BSDE formulation
for this purpose (but see Remark \ref{rkjointCharAnyway}).
We first have to clarify what we mean by a solution of the BSDE. We
consider $R$ and $A$ as given.
Since the finite variation part in the BSDE is predictable, a solution
will certainly be a special semimartingale. If $\ell$ is any special
semimartingale, there exists a unique orthogonal decomposition (\cite
{JacodShiryaev03}, III.4.24),
%
%
\begin{equation}\label{eqGeneralGKWdecompEll}
\ell= \ell_0 + A^\ell+ \varphi^\ell\sint R^c + W^\ell\ast(\mu
^R-\nu^R) + N^\ell,
\end{equation}
using the same notation as in (\ref{eqGeneralGKWdecompL}).
These processes are essentially unique, and so it suffices to consider
the left-hand side of the BSDE for the notion of a solution.
(In BSDE theory, a solution would be, at least, a quadruple.) %
We define the random function $g^\ell$ as in Corollary \ref
{coBellmanBsdeGen}, with $L$ replaced by $\ell$. Since $\ell$ is
special, we have
%
%
\begin{equation}\label{eqFRellIntegrates}
\int_{\R^d\times\R} (|x|^2+|x'|^2)\wedge(1+|x'|) F^{R,\ell
}(d(x,x'))<\infty,
\end{equation}
and the arguments from
Lemma \ref{lePropertiesOfI} show that $g^\ell$ is well defined on
$\sC
^0$ with values in $\R\cup\{\sgn(p) \infty\}$.
Hence, we can consider (formally at first) the BSDE (\ref
{eqBellmanBsdeGenForL}) with $L$ replaced by $\ell$, that is,
%
%
\begin{eqnarray}\label{eqBellmanBsdeGenForEll}
\ell&=& \ell_0 -p U^*(\ell_-)\sint\mu- p \max_{y\in\sC\cap\sC^0}
g^\ell(y)\sint A + \varphi^\ell\sint R^c\nonumber\\[-8pt]\\[-8pt]
&&{} + W^\ell\ast(\mu^R-\nu
^R) +
N^\ell\nonumber
\end{eqnarray}
with terminal condition $\ell_T=D_T$.
\begin{Def}\label{desolBellman}
A c\`adl\`ag special semimartingale $\ell$ is called a \textit{solution
of the Bellman equation} (\ref{eqBellmanBsdeGenForEll}) if:
\begin{itemize}
\item$\ell,\ell_->0$,
\item there exists a $\sC\cap\sC^{0,*}$-valued process $\cpi\in L(R)$
such that
\[
g^\ell(\cpi)=\sup_{\sC\cap\sC^{0}}g^\ell<\infty,
\]
\item$\ell$ and the processes from (\ref{eqGeneralGKWdecompEll})
satisfy (\ref{eqBellmanBsdeGenForEll}) with $\ell_T=D_T$.
\end{itemize}
Moreover, we define $\ckappa:=(D/\ell)^\beta$, where $\beta
=(1-p)^{-1}$. We call $(\cpi,\ckappa)$ the strategy associated with
$\ell$,
and for brevity, we also call $(\ell,\cpi,\ckappa)$ a solution.
\end{Def}

If the process $\cpi$ is not unique, we choose and fix one.
The assumption $\ell>0$ excludes pathological cases where $\ell$ jumps
to zero and becomes positive immediately afterwards, and thereby ensures
that $\ckappa$ is admissible. More precisely, the following holds.
\begin{Remark}\label{rksolutionBellman}
Let $(\ell,\cpi,\ckappa)$ be a solution of the Bellman equation.
\begin{longlist}
\item$(\cpi,\ckappa)\in\cA^{fE}$.
\item$\sup_{\sC\cap\sC^{0}}g^\ell$ is a predictable,
$A$-integrable process.
\item If $p\in(0,1)$, $g^\ell$ is finite on $\sC\cap\sC^{0}$.
\item The condition $\ell>0$ is automatically satisfied if either \textup{(a)} $p\in
(0,1)$ or if \textup{(b)}~$p<0$ and there is no intermediate consumption and
Assumptions~\ref{asbellmanSection} are satisfied.
\end{longlist}
\end{Remark}
\begin{pf}
(i) We have $\int_0^T \ckappa_s \mu(ds)<\infty$ $P$-a.s. since the
paths of $\ell$ are bounded away from zero.
Moreover, $\int_0^T U_t(\ckappa_t X_t(\cpi,\ckappa)) \mu
(dt)<\infty$
as in the proof of (\ref{eqoptPropConsGen}) (stated after Lemma \ref
{leZdriftrateNotPositive}). This shows $(\cpi,\ckappa)\in\cA^f$. The
fact that $(\cpi,\ckappa)\in\cA^{fE}$ is contained in the proof of
Lemma \ref{leBellmanMinimal} below.

\mbox{}\hphantom{i}(ii) We have $0=g^\ell(0)\leq\sup_{\sC\cap\sC^{0}}g^\ell=g^\ell
(\cpi
)$. Hence, $\sup_{\sC\cap\sC^{0}}g^\ell\sint A$ is well defined,
and it
is finite because otherwise (\ref{eqBellmanBsdeGenForEll}) could not hold.

(iii) Note that $p>0$ implies $g^\ell>-\infty$ by its definition
and (\ref{eqFRellIntegrates}), while $g^\ell<\infty$ by assumption.

\mbox{}\hspace*{1pt}(iv) If $p>0$, (\ref{eqBellmanBsdeGenForEll}) states that $A^\ell$ is
decreasing. As $\ell_->0$ implies $\ell\geq0$,
$\ell$~is a supermartingale by Lemma \ref{lenonnegDriftRateEquivalences}. %
Since $\ell_T=D_T>0$, the minimum principle for nonnegative
supermartingales shows $\ell>0$.
Under (b) the assertion is a~consequence of Theorem \ref
{thBellmanMinimal} below (which shows $\ell\geq L>0$) upon noting that
the condition $\ell>0$ is not used in its proof when there is no
intermediate consumption.
\end{pf}

It may seem debatable to make existence of the maximizer $\cpi$ part of
the definition of a solution. However, associating a control with the
solution is crucial for the following theory. Some justification is
given by the following result for the continuous case (where
$\sC^{0,*}=\R^d$).
\begin{Prop}\label{prMeasSelectionContCase}
Let $\ell$ be any c\`adl\`ag special semimartingale such that $\ell
,\ell_->0$. Under Assumptions \ref{ascontCase}, \textup{(C1)} and \textup{(C2)},
there exists a $\sC\cap\sC^{0,*}$-valued predictable process $\cpi$
such that $g^\ell(\cpi)=\sup_{\sC\cap\sC^{0}}g^\ell<\infty$,
and any
such process is $R$-integrable.
\end{Prop}
\begin{pf}
As $g^\ell$ is analogous to (\ref{eqDefOfgContinuous}), it is
continuous and its supremum over $\R^d$ is finite.
By continuity of $R$ and the structure condition, $\pi\in L(R)$ if and
only if $\int_0^T \pi^\top \,d\langle M \rangle\pi=\int_0^T |\sigma
^\top\pi|^2\,
dA<\infty$ $P$-a.s.

Assume first that $\sC$ is compact, then Lemma \ref{lemeasMaxThm}
yields a measurable selector $\pi$ for $\argmax_\sC g$.
As in the proof of Corollary \ref{coBellmanBsdeCont},
$\sigma^\top\pi\in\Pi^{\sigma^\top\sC} \sigma^\top\psi$ holds for
$\psi:=\beta(\lambda+\frac{\varphi^\ell}{\ell_{-}})$, which
satisfies $\int_0^T |\sigma^\top\psi|^2 \,dA<\infty$ by definition of~%
$\lambda$ and $\varphi^\ell$.
We note that
$|\sigma^\top\pi|\leq|\sigma^\top\psi| + |\sigma^\top\pi-
\sigma
^\top\psi| \leq2 |\sigma^\top\psi|$
due to the definition of the projection and $0\in\sC$.

In the general case we approximate $\sC$ by a sequence of compact
constraints $\sC^n:=\sC\cap\{x\in\R^d\dvtx|x|\leq n\}$, each of which
yields a selector $\pi^n$ for $\argmax_{\sC^n}g$.
By the above, $|\sigma^\top\pi^n|\leq2 |\sigma^\top\psi|$, so the
sequence $(\sigma^\top\pi^n)_n$ is bounded for fixed $(\omega,t)$. A
random index argument as in the proof of Lemma \ref{lemeasMaxThm}
yields a selector $\vartheta$ for a cluster point of this sequence. We
have $\vartheta\in\sigma^\top\sC$
by closedness of this set and we find a selector $\check{\pi}$ for
$((\sigma^\top)^{-1}\vartheta)\cap\sC$ using \cite
{Rockafellar76}, 1Q. We have $\cpi\in\argmax_{\sC}g$ as the sets
$\sC
^n$ increase to $\sC$, and
$\int_0^T |\sigma^\top\cpi|^2 \,dA \leq2 \int_0^T|\sigma^\top\psi
|^2 \,dA<\infty$ shows $\cpi\in L(R)$.
\end{pf}

Another example for the construction of $\cpi$ is given in \cite
{Nutz09c}, Section 5. In general, two ingredients are needed: existence
of a maximizer for fixed $(\omega,t)$
will typically require a compactness condition in the form of a
no-arbitrage assumption (in the previous proof, this is the structure
condition). Moreover, a measurable selection is required; here the
techniques from the \hyperref[app]{Appendices} may be useful.
\begin{Remark}\label{rkjointCharAnyway}
The BSDE formulation of the Bellman equation has the advantage that we
can choose $A$ based on $R$ and speak about the class of all solutions.
However, we do not want to write proofs in this cumbersome notation.
Once we fix a solution $\ell$ (and maybe $L$, and finitely many other
semimartingales), we can choose a new reference process $\tilde
{A}=A+A'$ (where $A'$ is increasing), with respect to which our
semimartingales admit differential characteristics; in particular we
can use the joint characteristics $(b^{R,\ell},c^{R,\ell},F^{R,\ell
};\tilde{A})$. As we change $A$, all drift rates change in that they
are multiplied by $d\tilde{A}/dA$, so any (in)equalities between them
are preserved. With this in mind, we shall use the joint
characteristics of $(R,\ell)$ in the sequel without further comment and
treat the two formulations of the Bellman equation as equivalent.
\end{Remark}

Our definition of a solution of the Bellman equation is loose in terms
of integrability assumptions. Even in the continuous case, it is
unclear ``how many'' solutions exist. The next result shows that we can
always identify $L$ by taking the smallest one, that is, $L\leq\ell$
for any solution~$\ell$.
\begin{Thm}\label{thBellmanMinimal}
Under Assumptions \ref{asbellmanSection}, the opportunity process $L$
is characterized as the minimal solution of the Bellman equation.
\end{Thm}
\begin{Remark}
As a consequence, the Bellman equation has a bounded solution if and
only if the opportunity process is bounded
(and similarly for other integrability properties). In conjunction
with \cite{Nutz09a}, Section 4.2, this yields examples of quadratic
BSDEs which have bounded terminal value (for~$D_T$ bounded), but no
bounded solution.
\end{Remark}

The proof of Theorem \ref{thBellmanMinimal} is based on the following
result; it is the fundamental property of any Bellman equation.\vadjust{\goodbreak}

\begin{Prop}\label{prBellmanSolSupermartingale}
Let $(\ell,\cpi,\ckappa)$ be a solution of the Bellman equation.
For any $(\pi,\kappa)\in\cA^f$,
%
%
\begin{equation}\label{eqZprocessinVerif}
Z(\pi,\kappa):=\ell\frac{1}{p}(X(\pi,\kappa))^p+\int U_s
(\kappa_sX_s(\pi,\kappa)) \mu(ds)
\end{equation}
is a semimartingale with nonpositive drift rate. Moreover, $Z(\cpi
,\ckappa)$ is a local martingale.\vspace*{-2pt}
\end{Prop}
\begin{pf}
Let $(\pi,\kappa)\in\cA^f$. Note that $Z:=Z(\pi,\kappa)$ satisfies
$\sgn(p)Z\geq0$, hence has a well-defined drift rate $a^Z$ by
Remark \ref{rkDriftRateNonnegSM}. The drift rate can be calculated as in
Lemma \ref{leCompensatorJointChar}: if $f^\ell$ is defined similarly to
the function $f$ in that lemma but with $L$ replaced by $\ell$, then
\begin{eqnarray*}%
a^{Z} &=& X(\pi,\kappa)^{p}_- \biggl\{ p^{-1} a^\ell+ f^\ell(\kappa)
\,\frac{d\mu}{dA} + g^\ell(\pi)\biggr\}\\[-2pt]
&=& X(\pi,\kappa)^{p}_- \biggl\{ \bigl(f^\ell(\kappa)-f^\ell(\ckappa)
\bigr) \,\frac{d\mu}{dA} + g^\ell(\pi) - g^\ell(\cpi)\biggr\}.
\end{eqnarray*}
This is nonpositive because $\ckappa$ and $\cpi$ maximize $f^\ell$ and
$g^\ell$.
For the special case $(\pi,\kappa):=(\cpi,\ckappa)$ we have $a^Z=0$
and so $Z$ is a $\sigma$-martingale, thus a local martingale as $\sgn
(p)Z\geq0$.\vspace*{-2pt}
\end{pf}
\begin{Remark}
In Proposition \ref{prBellmanSolSupermartingale}, ``semimartingale
with nonpositive drift rate'' can be replaced by ``$\sigma
$-supermartingale'' if $g^\ell$ is finite on $\sC\cap\sC^0$.\vspace*{-2pt}
\end{Remark}

Theorem \ref{thBellmanMinimal} follows from the next lemma (which is
actually stronger).
We recall that for $p<0$ the opportunity process $L$ can be defined
without further assumptions.\vspace*{-2pt}
\begin{Lemma}\label{leBellmanMinimal}
Let $\ell$ be a solution of the Bellman equation.
If $p<0$, then \mbox{$L\leq\ell$}. For $p\in(0,1)$, the same holds if
(\ref{eqPrimalProblemFinite}) is satisfied and
there exists an optimal strategy.\vspace*{-2pt}
\end{Lemma}
\begin{pf}%
Let $(\ell,\cpi,\ckappa)$ be a solution and define $Z(\pi,\kappa)$ as
in (\ref{eqZprocessinVerif}).

\textit{Case $p<0$}:
we choose $(\pi,\kappa):=(\cpi,\ckappa)$.
As $Z(\cpi,\ckappa)$ is a negative local martingale by Proposition
\ref
{prBellmanSolSupermartingale}, it is a submartingale.
In particular, $E[Z_T(\cpi,\ckappa)]>-\infty$, and using $L_T=D_T$,
this is the statement that the expected utility is finite, that is,
$(\cpi,\ckappa)\in\cA^{fE}$---this completes the proof of Remark
\ref
{rksolutionBellman}(i).
Recall that $\mu^\circ=\mu+\delta_{\{T\}}$.
With $\cX:=X(\cpi,\ckappa)$ and $\cc:=\ckappa\cX$, and using
$\ell
_T=D_T=L_T$, we deduce
\begin{eqnarray*}
&&\ell_t\frac{1}{p} \cX_t^p + \int_0^t
U_s(\cc_s) \mu(ds)\\[-2pt]
&&\qquad=Z_t(\cpi,\ckappa) \leq E[Z_T(\cpi,\ckappa
)|\cF_t]\\[-2pt]
&&\qquad\leq\esssup_{\tilde{c}\in\cA(\cpi,\cc,t)} E\biggl[\int_t^T
U_s(\tilde
{c}_s) \mu^\circ(ds)\Big|\cF_t\biggr] + \int_0^t U_s(\cc_s) \mu(ds)\\[-2pt]
&&\qquad= L_t\frac{1}{p}\cX_t^p + \int_0^t U_s(\cc_s) \mu(ds),
\end{eqnarray*}
where the last equality holds by (\ref{eqOppProcIndep}).
As $\frac{1}{p}\check{X}_t^p<0$, we have $\ell_t\geq L_t$.

\textit{Case $p\in(0,1)$}:
We choose $(\pi,\kappa):=(\hpi,\hkappa)$ to be an optimal strategy.
Then $Z(\hpi,\hkappa)\geq0$ is a supermartingale by
Proposition \ref{prBellmanSolSupermartingale} and Lemma \ref
{lenonnegDriftRateEquivalences}(iii), and we obtain
\begin{eqnarray*}
\ell_t\frac{1}{p} \hX_t^p + \int_0^t
U_s(\hc_s) \mu(ds)&=&Z_t(\hpi,\hkappa) \geq E[Z_T(\hpi,\hkappa
)|\cF_t]\\
&=&E\biggl[\int_0^T U_s(\hc_s) \mu^\circ(ds)\Big|\cF_t\biggr]\\
&=&L_t\frac
{1}{p} \hX_t^p + \int_0^t U_s(\hc_s) \mu(ds)
\end{eqnarray*}
by the optimality of $(\hpi,\hkappa)$ and (\ref{eqOppProcIndep}).
More precisely, we have used the fact that $(\hpi,\hkappa)$ is also
conditionally optimal (see \cite{Nutz09a}, Remark 3.3).
As $\frac{1}{p}\hX_{t}^p>0$, we conclude $\ell_t\geq L_t$.
\end{pf}

\section{Verification}\label{severification}

Suppose that we have found a solution of the Bellman equation; then we
want to know whether it is the opportunity process and whether the
associated strategy is optimal. In applications, it might not be clear
a priori that an optimal strategy exists or even that the utility
maximization problem is finite. Therefore, we stress that
in this section these properties are not assumed.
Also, we do not need the assumptions on~$\sC$ made in Section \ref
{seConstraintsDegeneracies}---they are not necessary because we start
with a~given solution.

Generally speaking, verification involves the candidate for an optimal
control, $(\cpi,\ckappa)$ in our case, and all
the competing ones. It is often very difficult to check a condition
involving all these controls, so it is desirable
to have a~verification theorem whose assumptions involve only $(\cpi
,\ckappa)$.

We present two verification approaches. The first one is via the value
process and is classical for general dynamic programming: it uses
little structure of the given problem. For $p\in(0,1)$, it yields
the desired result. However, in a general setting, this is not the case
for $p<0$.
The second approach uses the concavity of the utility function. To
fully exploit this and make the verification conditions necessary, we
will assume that $\sC$ is convex. In this case, we shall obtain the
desired verification theorem for all values of $p$.

\subsection{Verification via the value process}

The basis of this approach is the following simple\vadjust{\goodbreak} result; we state it
separately for better comparison with
Lemma \ref{leverificationGen} below. In the entire section, $Z(\pi
,\kappa)$ is defined by (\ref{eqZprocessinVerif})
whenever~$\ell$ is given.\vspace*{-2pt}
\begin{Lemma}\label{leverificationBasisDirect}
Let $\ell$ be any positive c\`adl\`ag semimartingale with $\ell_T=D_T$
and let $(\cpi,\ckappa)\in\cA$.
Assume that for all $(\pi,\kappa)\in\cA^{fE}$, the process $Z(\pi
,\kappa)$ is a supermartingale. Then $Z(\cpi,\ckappa)$ is a martingale
if and only if
(\ref{eqPrimalProblemFinite}) holds and $(\cpi,\ckappa)$ is optimal
and $\ell=L$.\vspace*{-2pt}
\end{Lemma}
\begin{pf}
``$\Rightarrow$'': Recall that $Z_0(\pi,\kappa)=\ell_0 \frac
{1}{p}x_0^p$ does not depend on $(\pi,\kappa)$ and that
$E[Z_T(\pi,\kappa)]=E[\int_0^T U_t(\kappa_t(X_t(\pi,\kappa))) \mu
^\circ(dt)]$ is the expected utility corresponding to $(\pi,\kappa)$.
With $\cX:=X(\cpi,\ckappa)$, the (super)martingale condition implies that
$E[\int_0^T U_t(\ckappa_t\cX_t) \mu^\circ(dt)]\geq E[\int_0^T
U_t(\kappa_t X_t(\pi,\kappa)) \mu^\circ(dt)]$ for all $(\pi
,\kappa)\in
\cA^{fE}$. Since for $(\pi,\kappa)\in\cA\setminus\cA^{fE}$ the expected
utility is $-\infty$, this\vspace*{1pt} shows that $(\cpi,\ckappa)$ is optimal with
$E[Z_T(\cpi,\ckappa)]=Z_0(\cpi,\ckappa)=\ell_0 \frac
{1}{p}x_0^p<\infty
$. In particular,
the opportunity process $L$ is well defined. By Proposition \ref
{prmartOptPrincipleForL},
$L\frac{1}{p}\cX^p + \int U_s(\cc_s) \mu(ds)$ is a martingale, and
as its terminal value equals~$Z_T(\cpi,\ckappa)$,
we deduce $\ell=L$ by comparison with (\ref{eqZprocessinVerif}), using
$\cX>0$.

The converse is contained in Proposition
\ref{prmartOptPrincipleForL}.\vspace*{-2pt}
\end{pf}

We can now state our first verification theorem.\vspace*{-2pt}
\begin{Thm}\label{thDirectVerification}
Let $(\ell,\cpi,\ckappa)$ be a solution of the Bellman equation.
\begin{enumerate}[(ii)]
\item[(i)] If $p\in(0,1)$, the following are equivalent:
\begin{enumerate}[(a)]
\item[(a)]$Z(\cpi,\ckappa)$ is of class (D),
\item[(b)]$Z(\cpi,\ckappa)$ is a martingale,
\item[(c)](\ref{eqPrimalProblemFinite}) holds and $(\cpi,\ckappa)$ is
optimal and $\ell=L$.
\end{enumerate}
\item[(ii)] If $p<0$, the following are equivalent:
\begin{enumerate}[(a)]
\item[(a)]$Z(\pi,\kappa)$ is of class (D) for all $(\pi,\kappa)\in\cA^{fE}$,
\item[(b)]$Z(\pi,\kappa)$ is a supermartingale for all $(\pi,\kappa)\in
\cA^{fE}$,
\item[(c)]$(\cpi,\ckappa)$ is optimal and $\ell=L$.\vspace*{-2pt}
\end{enumerate}
\end{enumerate}
\end{Thm}
\begin{pf}
When $p>0$ and $(\pi,\kappa)\in\cA^f$, $Z(\pi,\kappa)$ is
positive and
$a^{Z(\pi,\kappa)}\leq0$ by Proposition \ref
{prBellmanSolSupermartingale}, hence, $Z(\pi,\kappa)$ is a
supermartingale according to Lem\-ma~\ref{lenonnegDriftRateEquivalences}.
By Proposition \ref{prBellmanSolSupermartingale}, $Z(\cpi,\ckappa)$
is a local martingale, so it is a martingale if and only if it is of
class (D).
Lemma \ref{leverificationBasisDirect} implies the result.

If $p<0$, $Z(\pi,\kappa)$ is negative.
Thus the local martingale $Z(\cpi,\ckappa)$ is a submartingale, and a
martingale if and only if it is also a supermartingale.
Note that a class (D) semimartingale with nonpositive drift rate is a
supermartingale. Conversely, any negative supermartingale $Z$ is of
class (D) due to the bounds $0\geq Z \geq E[Z_T|\F]$.
Lemma \ref{leverificationBasisDirect} implies the result after noting
that if $\ell=L$, then Proposition \ref{prmartOptPrincipleForL}
yields (b).\vspace*{-2pt}
\end{pf}

Theorem \ref{thDirectVerification} is ``as good as it gets'' for $p>0$,
but as announced, the result for $p<0$ is not satisfactory.
In particular settings, this can be improved.\vadjust{\goodbreak}
\begin{Remark}[($p<0$)]\label{rkdirectVerificationBoundedL}
\textup{(i)} Assume we know a priori that \textup{if} there is an optimal strategy
$(\hpi,\hkappa)\in\cA$, then
\[
(\hpi,\hkappa)\in\cA^{(D)}:=\{(\pi,\kappa)\in\cA\dvtx X(\pi
,\kappa
)^p \mbox{ is of class (D)}\}.
\]
In this case we can reduce our optimization problem
to the class $\cA^{(D)}$. If, in addition, $\ell$ is bounded (which is
not a strong assumption when $p<0$),
the class (D) condition in Theorem \ref{thDirectVerification}\textup{(ii)} is
automatically satisfied for $(\pi,\kappa)\in\cA^{(D)}$.
The verification then reduces to checking that $(\cpi,\ckappa)\in\cA^{(D)}$.

\textup{(ii)} How can we establish the condition needed for \textup{(i)?}
One possibility is to show that $L$ is uniformly bounded away from
zero; then the condition follows (see the argument in the next proof).
Of course, $L$ is not known when we try to apply this.
However, \cite{Nutz09a}, Section 4.2, gives verifiable conditions for
$L$ to be (bounded and) bounded away from zero. They are stated
for the unconstrained case $\sC=\R^d$, but can be used nevertheless:
if $L^{\R^d}$ is the opportunity process corresponding to $\sC=\R^d$,
the actual $L$ satisfies $L\geq L^{\R^d}$ because the supremum
in (\ref{eqOppProcIndep}) is taken over a smaller set in the
constrained case.
\end{Remark}

In the situation where $\ell$ and $L^{-1}$ are bounded, we can also use
the following result. Note also its use in Remark \ref
{rkAfterBellmanThm}(ii) and recall that $1/0:=\infty$.

\begin{Cor}\label{codirectVerificationBoundedL}
Let $p<0$ and let $(\ell,\cpi,\ckappa)$ be a solution of the Bellman
equation.
Let $L$ be the opportunity process and assume that $\ell/L$ is
uniformly bounded.
Then $(\cpi,\ckappa)$ is optimal and $\ell=L$.
\end{Cor}
\begin{pf}
Fix arbitrary $(\pi,\kappa)\in\cA^{fE}$ and let $X=X(\pi,\kappa)$.
The process $L\frac{1}{p}(X(\pi,\kappa))^p+\int U_s(\kappa
_sX_s) \mu(ds)$
is a negative supermartingale by Proposition~\ref
{prmartOptPrincipleForL}, hence, of class (D). Since
$\int U_s(\kappa_sX_s) \mu(ds)$ is\vspace*{1pt} decreasing and its terminal value
is integrable (definition of $\cA^{fE}$),
$L\frac{1}{p}X^p$ is also of class (D). The assumption yields that
$\ell\frac{1}{p}X^p$ is of class (D), and then so is $Z(\pi,\kappa)$.
\end{pf}

As bounded solutions are of special interest in BSDE theory, let us
note the following consequence.
\begin{Cor}\label{coRevHolderAndBellmanBSDE}
Let $p<0$. Under Assumptions \ref{asbellmanSection} the following are
equivalent:
\begin{longlist}
\item$L$ is bounded and bounded away from zero;
\item there exists a unique bounded solution of the Bellman equation,
and this solution is
bounded away from zero.
\end{longlist}
\end{Cor}

One can note that in the setting of \cite{Nutz09a}, Section 4.2, these
conditions are further equivalent to a reverse H\"older inequality for
the market model.

We give an illustration of Theorem \ref{thDirectVerification} also for
the case $p\in(0,1)$. Thus far, we have considered only the given
exponent $p$
and assumed\vadjust{\goodbreak} (\ref{eqPrimalProblemFinite}). In many situations, there
will exist some $p_0\in(p,1)$ such that, if we consider
the exponent $p_0$ instead of $p$, the utility maximization problem is
still finite. Note that by Jensen's inequality
this is a stronger assumption.
We define for $q_0\geq1$ the class of semimartingales $\ell$ bounded in
$L^{q_0}(P)$,
\[
\mathbf{B}(q_0):=\Bigl\{\ell\dvtx\sup_\tau\|\ell_\tau\|
_{L^{q_0}(P)}<\infty\Bigr\},
\]
where the supremum ranges over all stopping times $\tau$.
\begin{Cor}\label{coVerifBoundedinLq}
Let $p\in(0,1)$ and let there be a constant $k_1>0$ such that $D\geq k_1$.
Assume that the utility maximization problem is finite for some
$p_0\in(p,1)$ and
let $q_0\geq1$ be such that $q_0>p_0/(p_0-p)$.
If $(\ell,\cpi,\ckappa)$ is a solution of the Bellman equation (for $p)$
with $\ell\in\mathbf{B}(q_0)$, then $\ell=L$ and $(\cpi,\ckappa)$
is optimal.
\end{Cor}
\begin{pf}
Let $\ell\in\mathbf{B}(q_0)$ be a solution, $(\cpi,\ckappa)$ the
associated strategy and let $\cX=X(\cpi,\ckappa)$.
By Theorem \ref{thDirectVerification} and an argument as in the
previous proof, it suffices to show that $\ell\cX^p$ is of class (D).
Let $\delta>1$ be such that $\delta/q_0+\delta p/p_0=1$. For every
stopping time $\tau$, H\"older's inequality yields
\[
E[(\ell_{\tau} \cX^p_{\tau})^\delta]=E[(\ell_{\tau
}^{q_0})^{\delta/q_0}
(\cX_{\tau}^{p_0})^{\delta p/p_0}]\leq E[\ell_{\tau}^{q_0}]^{\delta
/q_0} E[\cX_{\tau}^{p_0}]^{\delta p/p_0}.
\]
We show that this is bounded uniformly in $\tau$; then $\{\ell_{\tau}
\cX^p_{\tau}\dvtx\tau\mbox{ stopping time}\}$ is boun\-ded in
$L^\delta
(P)$ and hence uniformly integrable. Indeed, $E[\ell_{\tau}^{q_0}]$ is
boun\-ded by assumption. The
set of wealth processes corresponding to admissible strategies is
stable under stopping. Therefore,
$E[D_T\frac{1}{p_0}\cX_{\tau}^{p_0}]\leq u^{(p_0)}(x_0)$, the value
function for the utility maximization problem with exponent $p_0$. The
result follows as $D_T\geq k_1$.
\end{pf}
\begin{Remark}
In \cite{Nutz09a}, Example 4.6, we give a condition which implies that
the utility maximization problem is finite for \textup{all} $p_0\in(0,1)$.
Conversely, given such a $p_0\in(p,1)$, one can show that $L\in
\mathbf{B}(p_0/p)$ if $D$ is uniformly bounded from above (see \cite
{Nutz09d}, Corollary 4.2). %
\end{Remark}
\begin{Example}\label{exHuImkellerMuller}
We apply our results in an It\^o model with bounded mean--variance
tradeoff process together with an existence result for BSDEs. For the
case of utility from terminal wealth only, we retrieve (a minor
generalization of) the pioneering result of \cite{HuImkellerMuller05},
Section 3; the case with intermediate consumption is new.
Let $W$ be an $m$-dimensional standard Brownian motion ($m\geq d$) and
assume that $\F$ is generated by $W$. We consider
\[
dR_t=b_t \,dt + \sigma_t \,dW_t,
\]
where $b$ is predictable $\R^d$-valued and $\sigma$ is predictable
$\R^{d\times m}$-valued with everywhere full rank; moreover, we consider
constraints $\sC$ satisfying (C1) and (C2). We are in the situation of
Assumptions \ref{secontCase} with $dM=\sigma \,dW$ and $\lambda=
(\sigma\sigma^\top)^{-1}b$. The process $\theta:=\sigma^\top
\lambda$
is called \textit{market price of risk}.
We assume that there are constants $k_i>0$ such that
\[
0<k_1\leq D\leq k_2 \quad\mbox{and}\quad \int_0^T |\theta_s|^2 \,ds\leq k_3.
\]
The latter condition is called \textit{bounded mean--variance tradeoff}.
We remark that $dQ/ dP=\cE(-\lambda\sint M)_T=\cE(-\theta\sint W)_T$
defines a local martingale measure for~$\cE(R)$.
By \cite{Nutz09a}, Section 4.2, the utility maximization problem is
finite for all $p$ and the opportunity process
$L$ is bounded and bounded away from zero. It is continuous due to
Remark~\ref{rkLcontAndBSDEAppl}(i).

As suggested above, we write the Bellman BSDE for $Y:=\log(L)$ rather
than $L$ in this setting. If
$Y= A^Y + \varphi^Y\sint M + N^Y$ is the Kunita--Watanabe
decomposition, we write $Z:=\sigma^\top\varphi^Y$ and choose $Z^\bot$
such that \mbox{$Z^\bot\sint W=N^Y$} by Brownian representation.\vspace*{1pt} The
orthogonality of the decomposition implies $\sigma^\top Z^\bot=0$ and
$Z^\top Z^\bot=0$.
We write $\delta=1$ if there is intermediate consumption and $\delta
=0$ otherwise. Then It\^o's formula and Corollary \ref
{coBellmanBsdeCont} (with $A_t:=t$) yield the BSDE
%
%
\begin{equation}\label{eqHuImkellerMullerBSDE}
dY= f(Y,Z,Z^\bot) \,dt+ (Z+Z^\bot) \,dW;\qquad Y_T=\log(D_T)
\end{equation}
with
\begin{eqnarray*}
f(Y,Z,Z^\bot)&=& \frac{1}{2}p(1-p) d^2_{\sigma^\top\sC}\bigl(\beta
(\theta+ Z)\bigr)+\frac{q}{2} |\theta+ Z|^2 \\
&&{}+\delta(p-1)D^\beta\exp\bigl((q-1)Y\bigr) - \frac{1}{2}(|Z|^2+|Z^\bot|^2).
\end{eqnarray*}
Here $\beta=(1-p)^{-1}$ and $q=p/(p-1)$; the dependence on $(\omega,t)$
is suppressed in the notation. Using the orthogonality relations and
$p(1-p)\beta^2=-q$,\vspace*{1pt}
one can check that $f(Y,Z,Z^\bot)=f(Y,Z+Z^\bot,0)=:f(Y,\tZ)$, where
$\tZ:=Z+Z^\bot$.
As $0\in\sC$, we have
$d^2_{\sigma^\top\sC}(x)\leq|x|^2$. Hence,\vspace*{1pt} there exist a constant $C>0$
and an increasing continuous function $\phi$ such that
\[
|f(y,\tilde{z})| \leq C \bigl( |\theta|^2 +\phi(y)+|\tilde{z}|^2 \bigr).
\]
The following monotonicity property handles the exponential
nonlinearity caused by the consumption: as $p-1<0$ and $q-1<0$,
\[
-y [f(y,\tilde{z})-f(0,\tilde{z})]\leq0. %
\]
Thus we have Briand and Hu's \cite{BriandHu08}, Condition (A.1) after
noting that they call $-f$ what we call $f$, and \cite{BriandHu08},
Lemma 2 states the existence of a bounded solution $Y$ to the
BSDE (\ref{eqHuImkellerMullerBSDE}).
Let us check that $\ell:=\exp(Y)$ is the opportunity process. We
define an associated
strategy $(\cpi,\ckappa)$ by $\ckappa:=(D/\ell)^\beta$ and
Proposition \ref{prMeasSelectionContCase}; then we have a
solution $(\ell,\cpi,\ckappa)$ of the Bellman equation in the sense of
Definition \ref{desolBellman}.
For $p<0$ [$p\in(0,1)$], Corollary~\ref{codirectVerificationBoundedL}
(Corollary~\ref{coVerifBoundedinLq}) yields $\ell=L$ and the optimality
of $(\cpi,\ckappa)$. In fact, the same verification argument applies if
we replace $\cpi$ by any other predictable $\sC$-valued $\pi^*$ such that
$\sigma^\top\pi^* \in\Pi^{\sigma^\top\sC} \{\beta(\theta+Z)\}$;
recall from Proposition \ref{prMeasSelectionContCase} that $\pi^*\in
L(R)$ holds automatically.
To conclude: we have that
\[
L=\exp(Y)\mbox{ is the opportunity process},
\]
and the set of optimal strategies equals the set of all $(\pi
^*,\hkappa
)$ such that:
\begin{itemize}
\item$\hkappa=(D/L)^\beta$ $\mu^\circ$-a.e.,
\item$\pi^*$ is predictable, $\sC$-valued and $\sigma^\top\pi^*
\in
\Pi^{\sigma^\top\sC} \{\beta(\theta+Z)\}$ $P\otimes dt$-a.e.
\end{itemize}
One can remark that the previous arguments show $Y'=\log(L)$ whenever~$Y'$ is a solution of the BSDE (\ref{eqHuImkellerMullerBSDE}) which is
uniformly bounded from above. Hence, we have proved uniqueness
for (\ref{eqHuImkellerMullerBSDE}) in this class of solutions, which is
not immediate from BSDE theory.
One can also note that, in contrast to~\cite{HuImkellerMuller05}, we
did not use the theory of BMO martingales in this example. Finally,
we remark that the existence of an optimal strategy can also be
obtained by convex duality, under the additional assumption that $\sC$
is convex.
\end{Example}

We close this section with a formula intended for future applications.
\begin{Remark}\label{rkexponentialFormualsZ}
Let $(\ell,\cpi,\ckappa)$ be a solution of the Bellman equation.
Sometimes exponential
formulas can be used to verify that $Z(\cpi,\ckappa)$ is of class~(D).

Let $h$ be a predictable cut-off function such that $\cpi^\top h(x)$
is bounded,
for example, $h(x)=x1_{\{|x|\leq1\}\cap\{|\cpi^\top x|\leq1\}}$, and
define $\Psi$ to be the local martingale
\begin{eqnarray*}
&&\ell_-^{-1}\sint M^\ell+ p\cpi\sint R^c\\
&&\qquad{} + p\cpi^\top h(x)\ast(\mu
^R-\nu^R) + p(x'/\ell_-)\cpi^\top h(x) \ast(\mu^{R,\ell}-\nu
^{R,\ell})
\\
&&\qquad{}+ (1+x'/\ell_-)\{(1+\cpi^\top x)^p - 1 - p\cpi^\top h(x)\}
\ast(\mu^{R,\ell}-\nu^{R,\ell}).
\end{eqnarray*}
Then $\cE(\Psi)>0$, and if $\cE(\Psi)$ is of class (D), then
$Z(\cpi
,\ckappa)$ is also of class (D).
\end{Remark}

\begin{pf}
Let\vspace*{1pt} $Z=Z(\cpi,\ckappa)$. By a calculation as in the proof of Lem\-ma~\ref
{leCompensatorJointChar} and the local martingale condition from
Proposition \ref{prBellmanSolSupermartingale},
$(\frac{1}{p}\cX^p_-)^{-1}\sint Z =\ell_-\sint\Psi$.
Hence, $Z=Z_0\cE(\Psi)$ in the case without intermediate consumption.
For the general case, we have seen in the proof of Corollary \ref
{codirectVerificationBoundedL} that $Z$ is of class (D) whenever $\ell
\frac{1}{p}\cX^p$ is.
Writing the definition of $\ckappa$ as $\ckappa^{p-1}=\ell_-/D$ $\mu
$-a.e., we have
$
\ell\frac{1}{p}\cX^p=Z - \int\ckappa\ell_- \frac{1}{p}\cX^p\,
d\mu= (\ell_- \frac{1}{p}\cX^p_-)\sint(\Psi- \ckappa\sint\mu),
$
hence,
$
\ell\frac{1}{p}\cX^p = Z_0\cE(\Psi- \ckappa\sint\mu)=Z_0\cE
(\Psi
)\exp(- \ckappa\sint\mu).
$
It remains to note that $\exp(- \ckappa\sint\mu)\leq1$.
\end{pf}

\subsection{Verification via deflator}

The goal of this section is a verification theorem which involves
only
the candidate for the optimal strategy and holds\vadjust{\goodbreak} for general
semimartingale models. Our plan is as follows.
Let $(\ell,\cpi,\ckappa)$ be a solution of the Bellman\vspace*{1pt} equation and
assume for the moment that $\sC$ is convex. As the concave function
$g^\ell$ has a maximum at $\cpi$, the directional derivatives at
$\cpi$
in all directions should be nonpositive (if they can be defined).
A calculation will show that, at the level of processes, this yields a
supermartingale property which is well known from duality theory and
allows for verification.
In the case of nonconvex constraints, the directional derivatives need
not be defined in any sense. Nevertheless, the formally corresponding
quantities yield the expected result. To make the first-order
conditions necessary, we later specialize to convex $\sC$.
As in the previous section, we first state a basic result; it is
essentially classical.
\begin{Lemma}\label{leverificationGen}
Let $\ell$ be any positive c\`adl\`ag semimartingale with $\ell_T=D_T$.
Suppose there exists $(\cpi,\ckappa)\in\cA$ with $\ckappa=(D/\ell
)^\beta$ and let $\cX:=X(\cpi,\ckappa)$.
Assume $Y:=\ell\cX^{p-1}$ has the property that for all $(\pi,\kappa
)\in\cA$,
\[
\Gamma(\pi,\kappa):=X(\pi,\kappa)Y+\int\kappa_s X_s(\pi,\kappa
) Y_s
\mu(ds)
\]
is a supermartingale. Then $\Gamma(\cpi,\ckappa)$ is a martingale if
and only if
(\ref{eqPrimalProblemFinite}) holds and $(\cpi,\ckappa)$ is optimal
and $\ell=L$.
\end{Lemma}
\begin{pf}
``$\Rightarrow$'': let $(\pi,\kappa)\in\cA$ and denote $c=\kappa
X(\pi
,\kappa)$ and $\cc=\ckappa\cX$.
Note the partial derivative $\partial U(\cc)=D \ckappa^{p-1}\cX
^{p-1}=\ell\cX^{p-1}=Y$.
Concavity of $U$ implies $U(c)-U(\cc)\leq\partial U(\cc) (c-\cc)=Y
(c-\cc)$, hence,
\begin{eqnarray*}
&&
E\biggl[\int_0^T U_s(c_s) \mu^\circ(ds)\biggr]-E\biggl[\int_0^T U_s(\cc_s)
\mu^\circ(ds)\biggr]\\
&&\qquad\leq E\biggl[\int_0^T Y_s (c_s-\cc_s) \mu^\circ(ds) \biggr]\\
&&\qquad = E[\Gamma_T(\pi,\kappa)]-E[\Gamma_T(\cpi,\ckappa)].
\end{eqnarray*}
Let $\Gamma(\cpi,\ckappa)$ be a martingale; then $\Gamma_0(\pi
,\kappa
)=\Gamma_0(\cpi,\ckappa)$ and the supermartingale property imply that
the last line is nonpositive. As $(\pi,\kappa)$ was arbitrary, $(\cpi
,\ckappa)$ is optimal with expected utility
$E[\int_0^T U_s(\cc_s) \mu^\circ(ds)]=E[\frac{1}{p}\Gamma
_T(\cpi,\ckappa)]=\frac{1}{p}\Gamma_0(\cpi,\ckappa)=\frac{1}{p}x_0^p
\ell_0<\infty$.
The rest is as in the proof of Lemma \ref{leverificationBasisDirect}.
\end{pf}

The process $Y$ is a supermartingale deflator in the language of \cite
{KaratzasKardaras07}. We refer to \cite{Nutz09a} for
the connection of the opportunity process with convex duality, which in
fact suggests Lemma \ref{leverificationGen}.
Note that unlike $Z(\pi,\kappa)$ from the previous section, $\Gamma
(\pi
,\kappa)$ is positive for all values of $p$.

Our next goal is to link the supermartingale property to local
first-order conditions. Let $y,\cy\in\sC\cap\sC^0$ (we will plug in
$\cpi$
for $\cy$).
The formal directional derivative\vadjust{\goodbreak} of $g^\ell$ at $\cy$ in the direction
of $y$ is $(y-\cy)^\top\nabla g^\ell(\cy)=G^\ell(y,\cy)$, where,
by formal differentiation under the integral sign [cf. (\ref{eqDefOfg})],
%
%
\begin{eqnarray}\label{eqDirectDerivativeG}
G^\ell (y,\cy)
&:=& \ell_{-} (y-\cy)^\top\biggl( b^R + \frac{c^{R\ell}}{\ell_{-}}+
(p-1)c^R \cy\biggr)\nonumber\\
&&{} + \int_{\R^d\times\R} (y-\cy)^\top x' h(x)
F^{R,\ell
}(d(x,x')) \nonumber\\[-8pt]\\[-8pt]
&&{}+ \int_{\R^d\times\R} (\ell_-+x') \{(1+\cy^\top x)^{p-1}(y-\cy
)^\top x - (y-\cy)^\top h(x)\} \nonumber\\
&&\hphantom{{}+ \int_{\R^d\times\R}}{}\times F^{R,\ell}(d(x,x')).\nonumber
\end{eqnarray}
We take this expression as the definition of $G^\ell(y,\cy)$ whenever
the last integral is well defined [the first one is finite by (\ref
{eqFRellIntegrates})]. The differentiation
cannot be justified in general, but see the subsequent section.
\begin{Lemma}\label{leGwellDefandLsc}
Let $y\in\sC^0$ and $\cy\in\sC^{0,*}\cap\{g^\ell>-\infty\}$. Then
$G^\ell(y,\cy)$ is well defined with values in $(-\infty,\infty]$ and
$G^\ell(\cdot,\cy)$ is lower semicontinuous on~$\sC^0$.
\end{Lemma}
\begin{pf}
Writing $(y-\cy)^\top x=1+y^\top x - (1+\cy^\top x)$, we can express
$G^\ell(y,\cy)$ as
\begin{eqnarray*}%
&& \ell_{-} (y-\cy)^\top\biggl( b^R + \frac{c^{R\ell}}{\ell_{-}}+
(p-1)c^R \cy\biggr) + \int_{\R^d\times\R} (y-\cy)^\top x' h(x)
F^{R,\ell
}(d(x,x')) \\
&&\qquad{}+ \int_{\R^d\times\R} (\ell_-+x') \biggl\{\frac{1+y^\top x}{(1+\cy
^\top
x)^{1-p}} -1 -\bigl(y+(p-1)\cy\bigr)^\top h(x)\biggr\} \\
&&\qquad\hphantom{{}+ \int_{\R^d\times\R}}{}\times F^{R,\ell
}(d(x,x'))\\
&&\qquad{}- \int_{\R^d\times\R} (\ell_-+x') \{(1+\cy^\top x)^p -1 -p\cy
^\top h(x)\} F^{R,\ell}(d(x,x')).
\end{eqnarray*}
The first integral is finite and continuous in $y$ by (\ref
{eqFRellIntegrates}).
The last integral above occurs in the definition of $g^\ell(\cy)$
[cf. (\ref{eqDefOfg})] and it is finite if $g^\ell(\cy)>-\infty$ and
equals $+\infty$ otherwise. Finally,\vspace*{1pt} consider the second integral above
and call its integrand $\psi=\psi(y,\cy,x,x')$.
The Taylor expansion
$
\frac{1+y^\top x}{(1+\cy^\top x)^{1-p}} = 1 + (y+(p-1)\cy)^\top x +
\frac{(p-1)}{2}(2y + (p-2)\cy)^\top x x^\top\cy+ o(|x|^3)
$
shows that
$\int_{\{|x|+|x'|\leq1\}} \psi \,dF^{R,\ell}$ is well defined and
finite. It also shows that given a~compact $K\subset\R^d$, there is
$\eps>0$ such that $\int_{\{|x|+|x'|\leq\eps\}} \psi \,dF^{R,\ell}$ is
continuous in $y\in K$ (and also in $\cy\in K$).
The details are as in Lemma \ref{lePropertiesOfI}.
Moreover, for $y\in\sC^0$ we have the lower bound $\psi\geq(\ell
_-+x')\{-1 -(y+(p-1)\cy)^\top h(x)\}$, which is
$F^{R,\ell}$-integrable on $\{|x|+|x'|> \eps\}$ for any $\eps>0$,
again by (\ref{eqFRellIntegrates}).
The result now follows by Fatou's lemma.
\end{pf}

We can now connect the local first-order conditions for $g^\ell$ and
the global
supermartingale property: it turns out that the formal derivative
$G^\ell$ determines the sign of the drift rate of $\Gamma$,
cf. (\ref{eqDriftRateGamma}) below, which leads to the following
proposition. Here and in the sequel, we denote $\cX=X(\cpi,\ckappa)$.
\begin{Prop}\label{prsupermartAndGradient}
Let $(\ell,\cpi,\ckappa)$ be a solution of the Bellman equation and
$(\pi,\kappa)\in\cA$.
Then
$
\Gamma(\pi,\kappa):=\ell\cX^{p-1}X(\pi,\kappa)+\int\kappa_s
\ell_s\cX
^{p-1}_sX_s(\pi,\kappa) \mu(ds)
$
is a supermartingale (local martingale) if and only if $G^\ell(\pi
,\cpi)\leq0$ ($=0$).
\end{Prop}
\begin{pf}
Define $\bar{R}= R - (x-h(x))\ast\mu^R$ as in (\ref
{eqcanonicalRepR}). In the sequel, we abbreviate $\bar{\pi}:=(p-1)\cpi+\pi$ and
similarly $\bar{\kappa}:=(p-1)\ckappa+\kappa$.
We defer to Lemma \ref{leomittedCalc} a calculation showing
that $(\cX_-^{p-1} X_-(\pi,\kappa))^{-1}\sint(\ell\cX
^{p-1}X(\pi,\kappa))$ equals
\begin{eqnarray*}
&&\ell-\ell_0 + \ell_-\bar{\pi}\sint\bar{R} - \ell_- \bar
{\kappa}\sint\mu\\
&&\qquad{}+ \ell_-(p-1)\biggl(\frac{p-2}{2}\cpi+\pi\biggr)^\top c^R \cpi\sint A
+ \bar{\pi}^\top c^{R\ell}\sint A
+ \bar{\pi}^\top x'h(x) \ast\mu^{R,\ell}\\
&&\qquad{}+ (\ell_-+x')\{(1+\cpi^\top x)^{p-1}(1+\pi^\top x) - 1
- \bar{\pi}^\top h(x)\}\ast\mu^{R,\ell}.
\end{eqnarray*}
Here we use a predictable cut-off function $h$ such that $\bar{\pi
}^\top h(x)$ is bounded;
for example, $h(x)=x1_{\{|x|\leq1\}\cap\{|\bar{\pi}^\top x|\leq1\}}$.
Since $(\ell,\cpi,\ckappa)$ is a solution, the drift of $\ell$ is
\[
A^\ell= -pU^*(\ell_-)\sint\mu- p g^\ell(\cpi)\sint A = (p-1)\ell
_-\ckappa\sint\mu- p g^\ell(\cpi)\sint A.
\]
By Remark \ref{rkDriftRateNonnegSM}, $\Gamma:=\Gamma(\pi,\kappa)$
has a
well-defined drift rate $a^\Gamma$ with values in $(-\infty,\infty]$.
From the two formulas above and (\ref{eqcanonicalRepR}) we deduce
%
%
\begin{equation}\label{eqDriftRateGamma}
a^\Gamma= \cX_-^{p-1} X(\pi,\kappa)_- G^\ell(\pi,\cpi).
\end{equation}
Here $\cX_-^{p-1} X(\pi,\kappa)_- >0$ by admissibility.
If $\Gamma$ is a supermartingale, then $a^\Gamma\leq0$,
and the converse holds by Lemma \ref{lenonnegDriftRateEquivalences} in
view of $\Gamma\geq0$.
\end{pf}

We obtain our second verification theorem from Proposition \ref
{prsupermartAndGradient} and Lem\-ma~\ref{leverificationGen}.
\begin{Thm}\label{thVerificationDeflatorGen}
Let $(\ell,\cpi,\ckappa)$ be a solution of the Bellman equation.
Assume that $P\otimes A$-a.e.,
$G^\ell(y,\cpi)\in[-\infty,0]$ for all $y\in\sC\cap\sC^{0,*}$.
Then
\[
\Gamma(\cpi,\ckappa):=\ell\cX^p+\int\ckappa_s\ell_s \cX^p_s
\mu(ds)
\]
is a local martingale. It is a martingale if and only if (\ref
{eqPrimalProblemFinite}) holds and $(\cpi,\ckappa)$ is optimal and
$\ell
=L$ is the opportunity process.
\end{Thm}

If $\sC$ is not convex, one can imagine situations where
the directional derivative of $g^\ell$ at the maximum is positive,
that is,
the assumption on $G^\ell(y,\cpi)$ is sufficient but not
necessary. This changes in the subsequent section.

\subsubsection{The convex-constrained case}

We assume in this section that $\sC$ is convex; then $\sC\cap\sC^0$ is
also convex.
Our aim is to show that the nonnegativity condition on $G^\ell$ in
Theorem \ref{thVerificationDeflatorGen} is automatically satisfied in
this case.
We start with an elementary but crucial observation about
``differentiation under the integral sign.'' %
\begin{Lemma}\label{lediffUnderIntForConcave}
Consider two distinct points $y_0$ and $\cy$ in $\R^d$ and let $C=\{
\eta y_0 + (1-\eta)\cy\dvtx0\leq\eta\leq1\}$.
Let $\rho$ be a function on $\Sigma\times C$, where $\Sigma$ is some
Borel space with measure $\nu$, such that $x\mapsto\rho(x,y)$ is
$\nu
$-measurable, $\int\rho^+ (x,\cdot) \nu(dx)<\infty$ on~$C$, and
$y\mapsto\rho(x,y)$ is concave. In particular, the directional derivative
\[
D_{\cy,y}\rho(x,\cdot):= \lim_{\eps\to0+} \frac{\rho(x,\cy
+\eps
(y-\cy))-\rho(x,\cy)}{\eps}
\]
exists in $(-\infty,\infty]$ for all $y\in C$.
Let $\alpha$ be another concave function on $C$.

Define $\gamma(y):=\alpha(y)+\int\rho(x,y) \nu(dx)$ and assume that
$\gamma(y_0)>-\infty$ and that $\gamma(\cy)=\max_C \gamma<\infty$.
Then for all $y\in C$,
%
%
\begin{equation}\label{eqLemmadiffUnderIntForConcave}
D_{\cy,y}\gamma= D_{\cy,y}\alpha+ \int D_{\cy,y}\rho(x,\cdot)
\nu
(dx) \in(-\infty,0]
\end{equation}
and in particular $D_{\cy,y}\rho(x,\cdot)<\infty$ $\nu(dx)$-a.e.
\end{Lemma}
\begin{pf}
Note that $\gamma$ is concave, hence, we also have $\gamma>-\infty$ on
$C$. Let $v=(y-\cy)$ and $\eps>0$, then
$
\frac{\gamma(\cy+\eps v)-\gamma(\cy)}{\eps}=\frac{\alpha(\cy
+\eps
v)-\alpha(\cy)}{\eps} + \int\frac{\rho(x,\cy+\eps v)-\rho(x,\cy
)}{\eps
} \nu(dx).
$
By concavity, these quotients increase monotonically as $\eps
\downarrow0$, in particular their limits exist.
The left-hand side is nonpositive as $\cy$ is a maximum and monotone
convergence yields (\ref{eqLemmadiffUnderIntForConcave}).
\end{pf}

For completeness, let us mention that if $\gamma(y_0)=-\infty$, there
are examples where the left-hand side of (\ref
{eqLemmadiffUnderIntForConcave}) is $-\infty$ but the right-hand side is
finite; we shall deal with this case separately. We deduce the
following version of Theorem \ref{thVerificationDeflatorGen}; as
discussed, it involves only the control $(\cpi,\ckappa)$.
\begin{Thm}\label{thVerificationDeflatorConvex}
Let $(\ell,\cpi,\ckappa)$ be a solution of the Bellman equation and
assume that $\sC$ is convex.
Then
$\Gamma(\cpi,\ckappa):=\ell\cX^p+\int\ckappa_s\ell_s \cX^p_s
\mu(ds)$
is a local martingale.
It is a martingale if and only if (\ref{eqPrimalProblemFinite}) holds
and $(\cpi,\ckappa)$ is optimal and $\ell=L$.
\end{Thm}
\begin{pf}
To apply Theorem \ref{thVerificationDeflatorGen}, we have to check that
$G^\ell(y,\cpi)\in[-\infty,0]$ for $y\in\sC\cap\sC^{0,*}$.
Recall that $\cpi$ is a maximizer for $g^\ell$ and that $G^\ell$ was
defined by
differentiation under the integral sign. Lemma \ref
{lediffUnderIntForConcave} yields $G^\ell(y,\cpi)\leq0$ whenever
$y\in\{
g^\ell>-\infty\}$.
This ends the proof for $p\in(0,1)$ as $g^\ell$ is then finite.
If $p<0$, the definition of $g^\ell$ and Remark \ref{rkdiamondDef}
show that the set $\{g^\ell>-\infty\}$
contains the set $\bigcup_{\eta\in[0,1)} \eta(\sC\cap\sC^0)$ which,
in turn, is dense in $\sC\cap\sC^{0,*}$.
Hence, $\{g^\ell>-\infty\}$
is dense in $\sC\cap\sC^{0,*}$ and we obtain $G^\ell(y,\cpi)\in
[-\infty,0]$ for all $y\in\sC\cap\sC^{0,*}$ using the lower
semicontinuity from Lemma \ref{leGwellDefandLsc}.
\end{pf}
\begin{Remark}\label{rkexponentialFormualsGamma}
\textup{(i)} We note that $\Gamma(\cpi,\ckappa)=p Z(\cpi,\ckappa)$ if $Z$ is
defined as in~(\ref{eqZprocessinVerif}). In particular,
Remark \ref{rkexponentialFormualsZ} can be used also for $\Gamma(\cpi
,\ckappa)$.

\textup{(ii)} Muhle--Karbe \cite{MuhleKarbe09} considers certain
one-dimensional (unconstrained) affine models and introduces a
sufficient optimality condition in the form of an algebraic inequality
(see \cite{MuhleKarbe09}, Theorem 4.20(3)). This condition can be seen
as a special case of the statement that
$G^L(y,\cpi)\in[-\infty,0]$ for $y\in\sC^{0,*}$; in particular, we
have shown its necessity.
\end{Remark}

Of course, all our verification results can be seen as a uniqueness
result for the Bellman equation.
As an example, Theorem \ref{thVerificationDeflatorConvex} yields the
following corollary.
\begin{Cor}
If $\sC$ is convex, there is at most one solution of the Bellman
equation in the class of solutions $(\ell,\cpi,\ckappa)$ such that
$\Gamma(\cpi,\ckappa)$ is of class (D).
\end{Cor}

Similarly, one can give corollaries for the other results. We close
with a~comment concerning convex duality.
\begin{Remark}\label{rklossOfMass}
\textup{(i)} A major insight in \cite{KramkovSchachermayer99} was that the
``dual domain'' for utility maximization (here with $\sC=\R^d$) should
be a set of supermartingales rather than (local) martingales when the
price process has jumps. A~one-period example for $\log$-utility (\cite
{KramkovSchachermayer99}, Example 5.1$'$) showed that the supermartingale
solving the dual problem can indeed have nonvanishing drift. In that
example it is clear that this arises when the budget constraint becomes
binding. For general models and $\log$-utility, \cite{GollKallsen03}
comments on this phenomenon.
The calculations of this section yield an instructive ``local''
picture also for power utility.

Under Assumptions \ref{asbellmanSection}, the opportunity process $L$
and the optimal strategy $(\hpi,\hkappa)$ solve the Bellman equation.
Assume that $\sC$ is convex and let $\hX=X(\hpi,\hkappa)$.
Consider $\hY=L\hX^{p-1}$, which was the solution to the dual problem
in \cite{Nutz09a}.
We have shown that $\hY\cE(\pi\sint R)$ is a supermartingale for
every $\pi\in\cA$; that is, $\hY$ is a supermartingale deflator.
Choosing $\pi=0$, we see that~$\hY$ is itself a supermartingale, and
by (\ref{eqDriftRateGamma}) its drift rate satisfies
\[
a^{\hY}=\hX_-^{p-1} G^L(0,\hpi)=- \hX_-^{p-1} \hpi^\top\nabla
g(\hpi).
\]
Hence, $\hY$ is a local martingale if and only if $\hpi^\top\nabla
g(\hpi)=0$.
One can say that $-\hpi^\top\nabla g(\hpi)<0$ means that the
constraints are binding, whereas in an ``unconstrained'' case the
gradient of $g$ would vanish, that is, $\hY$ has nonvanishing drift
rate at a given $(\omega,t)$ whenever the constraints are binding. Even
if $\sC=\R^d$, we still have the budget constraint $\sC^0$ in the
maximization of $g$.
If, in addition, $R$ is continuous, $\sC^0=\R^d$ and we are truly in
an unconstrained situation. Then $\hY$ is a local martingale; indeed,
in the setting of Corollary \ref{coBellmanBsdeCont} we calculate
\[
\hY=y_0\cE\biggl(-\lambda\sint M + \frac{1}{L_{-}}\sint
N^L\biggr),\qquad
y_0:=L_0 x_0^{p-1}.
\]
Note how $N^L$, the martingale part of $L$ orthogonal to $R$, yields
the solution to the dual problem.

\textup{(ii)} From the proof of Proposition \ref{prsupermartAndGradient} we
have that the general formula for the local martingale part of $\hY$ is
\begin{eqnarray*}
M^{\hY}&=& \hX_-^{p-1} \sint\bigl( M^L + L_-(p-1)\hpi\sint
M^{\bar{R}}\\
&&\hphantom{\hX_-^{p-1} \sint\bigl(}\hspace*{0pt}{} + (p-1)\hpi^\top x'h(x) \ast(\mu^{R,L}-\nu^{R,L}) \\
&&\hphantom{\hX_-^{p-1} \sint\bigl(}{} + (L_-+x')\{(1+\hpi^\top x)^{p-1} - 1 - (p-1)\hpi^\top
h(x)\}\\
&&\hspace*{191.5pt}{}\ast(\mu^{R,L}-\nu^{R,L})\bigr).
\end{eqnarray*}
This is relevant in the problem of \textit{$q$-optimal equivalent
martingale measures}; cf. Goll and R\"uschendorf \cite
{GollRuschendorf01} for a general perspective. Let $u(x_0)<\infty$,
\mbox{$D\equiv1$}, \mbox{$\mu=0$}, $\sC=\R^d$, and assume that the set $\sM$ of
equivalent local martingale measures for $S=\cE(R)$ is nonempty.
Given $q=p/(p-1)\in(-\infty,0)\cup(0,1)$ conjugate to $p$,
$Q\in\sM$ is called \textit{$q$-optimal} if $E[-q^{-1}(dQ/dP)^q]$ is
finite and minimal over $\sM$. If $q<0$, that is, $p\in(0,1)$, then
$u(x_0)<\infty$ is equivalent to the existence of some $Q\in\sM$ such
that $E[-q^{-1}(dQ/dP)^q]<\infty$; moreover, Assumptions \ref
{asbellmanSection} are satisfied (see Kramkov and Schachermayer \cite
{KramkovSchachermayer99,KramkovSchachermayer03}).
Using \cite{KramkovSchachermayer99}, Theorem~2.2\textup{(iv)}, we conclude that:
\begin{longlist}[(a)]
\item[(a)] the $q$-optimal martingale measure exists if and only if
$a^{\hY}\equiv0$ and $M^{\hY}$ is a true martingale;
\item[(b)] in that case, $1+y_0^{-1}M^{\hY}$ is its $P$-density process.
\end{longlist}
This generalizes earlier results of \cite{GollRuschendorf01} as well as
of Grandits \cite{Grandits00}, Jeanblanc, Kl\"oppel and Miyahara \cite
{JeanblancKloppelMiyahara07} and Choulli and Stricker \cite{ChoulliStricker09}.
\end{Remark}

\begin{appendix}\label{app}
\section{\texorpdfstring{Proof of Lemma \lowercase{\protect\ref{lemeasmaxsequence}}: A measurable
maximizing~sequence}
{Proof of Lemma 3.8: A measurable maximizing~sequence}}
\label{seMeasSelection}

The main goal of this Appendix is to construct a measurable maximizing
sequence for the random function $g$; cf. Lemma \ref{lemeasmaxsequence}.
The entire section is under Assumptions\vadjust{\goodbreak} \ref{asbellmanSection}.
Before beginning the proof, we discuss the properties of $g$; recall that
%
%
\begin{eqnarray}\label{eqDefOfgRestated}
g(y)
&:=& L_{-}y^\top\biggl( b^R + \frac{c^{RL}}{L_{-}}+ \frac{(p-1)}{2}
c^R y \biggr)\nonumber\\
&&{} + \int_{\R^d\times\R} x' y^\top h(x) F^{R,L}(d(x,x'))
\nonumber\\[-8pt]\\[-8pt]
&&{}+ \int_{\R^d\times\R} (L_-+x') \{p^{-1}(1+y^\top x)^p - p^{-1} -
y^\top h(x)\} \nonumber\\
&&\hphantom{{} + \int_{\R^d\times\R}}{}\times F^{R,L}(d(x,x')).\nonumber
\end{eqnarray}
\begin{Lemma}\label{legFirstFactorPositive}
$L_-+x'$ is strictly positive $F^{L}(dx')$-a.e.
\end{Lemma}
\begin{pf}
We have
\begin{eqnarray*}
(P\otimes\nu^{L})\{L_-+x'\leq0\}
&=&E\bigl[1_{\{L_-+x'\leq0\}} \ast\nu^L_T\bigr]\\
&=&E\bigl[1_{\{L_-+x'\leq0\}} \ast\mu^L_T\bigr]\\
&=&E\biggl[\sum_{s\leq T} 1_{\{L_s\leq0\}}1_{\{\Delta L_s\neq0\}}\biggr],
\end{eqnarray*}
which vanishes as $L>0$ by Lemma \ref{leBoundsForL}.
\end{pf}

Fix $(\omega,t)$ and let $l:=L_{t-}(\omega)$.
Furthermore, let $F$ be any L\'evy measure on~$\R^{d+1}$ which is
equivalent to $F^{R,L}_t(\omega)$ and satisfies (\ref
{eqFRLintegrates}). Equivalence implies that $\sC^0_t(\omega),\sC
^{0,*}_t(\omega)$ and $\sN_t(\omega)$ are the same
if defined\vspace*{1pt} with respect to~$F$ instead of~$F^R$. Given $\eps>0$, let
\[
I^F_\eps(y):=\int_{\{|x|+|x'|\leq\eps\}} (l+x') \{
p^{-1}(1+y^\top x)^p - p^{-1} - y^\top h(x)\} F(d(x,x'))
\]
and
\[
I^F_{>\eps}(y):=\int_{\{|x|+|x'|> \eps\}} (l+x') \{
p^{-1}(1+y^\top x)^p - p^{-1} - y^\top h(x)\} F(d(x,x')),
\]
so that
\[
I^F(y):=I^F_\eps(y)+I^F_{>\eps}(y)
\]
is the last integral in (\ref{eqDefOfgRestated}) when
$F=F^{R,L}_t(\omega)$. We
know from the proof of Lem\-ma~\ref{leCompensatorJointChar} that
$I^{F^{R,L}}(\pi)$ is well defined and finite for any $\pi\in\cA^{fE}$
[of course, when $p>0$, this is essentially due to the assumption
(\ref{eqPrimalProblemFinite})].
For general $F$, $I^F$ has the following properties.
\begin{Lemma}\label{lePropertiesOfI} Consider a sequence $y_n\to
y_\infty$ in $\sC^0$.
\begin{longlist}
\item For any $y\in\sC^0$, the integral $I^F(y)$ is well defined in
$\R
\cup\{\sgn(p) \infty\}$.\vspace*{1pt}

\item For $\eps\leq(2 \sup_n |y_n|)^{-1}$ we have $I^F_\eps(y_n)\to
I^F_\eps(y_\infty)$.\vspace*{2pt}

\item If $p\in(0,1)$, then $I^F$ is l.s.c., that is, $\liminf_{n}
I^F(y_n)\geq I^F(y_\infty)$.

\item If $p<0$,\vspace*{1pt} then $I^F$ is u.s.c., that is, $\limsup_{n} I^F(y_n)\leq
I^F(y_\infty)$.
Moreover, $y\in\sC^0\setminus\sC^{0,\ast}$ implies $I^F(y)=-\infty$.
\end{longlist}
\end{Lemma}
\begin{pf}
The first item follows from the subsequent considerations.

(ii) We may assume that $h$ is the identity function on $\{|x|\leq\eps
\}$, then on this set
$
p^{-1}(1+y^\top x)^p - p^{-1} - y^\top h(x)=:\psi(z)|_{z=y^\top x},
$
where the function $\psi$ is smooth on $\{|z|\leq1/2\}\subseteq\R$
satisfying
\[
\psi(z)=p^{-1}(1+z)^p - p^{-1} - z = \frac{p-1}{2} z^2 + o(|z|^3),
\]
because $1+z$ is bounded away from $0$. Thus
$
\psi(z)=z^2\tilde{\psi}(z)
$
with a function~$\tilde{\psi}$ that is continuous and in particular
bounded on $\{|z|\leq1/2\}$.

As a L\'evy measure, $F$ integrates $(|x'|^2+|x|^2)$ on compacts; in particular,
$G(d(x,x')):=|x|^2 F(d(x,x'))$ defines a finite measure on $\{
|x|+|x'|\leq\eps\}$. Hence, $I^F_\eps(y)$ is well defined and finite
for $|y|\leq(2\eps)^{-1}$, and
dominated convergence shows that
$
I^F_\eps(y)=\int_{\{|x|+|x'|\leq\eps\}} (l+x') \tilde{\psi
}(y^\top
x) G(d(x,x'))
$
is continuous in $y$ on $\{|y|\leq(2\eps)^{-1}\}$.

(iii) For $|y|$ bounded by a constant $C$, the integrand in
$I^F$ is bounded from below by $C'+|x'|$ for some constant $C'$
depending on $y$ only through~$C$. We choose $\eps$ as before.
As $C'+|x'|$ is $F$-integrable on $\{|x|+|x'|> \eps\}$ by~(\ref
{eqFRLintegrates}), $I^F(y)$ is well defined in $\R\cup\{\infty\}$ and
l.s.c. by Fatou's lemma.

(iv) The first part follows as in (iii), now the
integrand is bounded from above by $C'+|x'|$.
If $y\in\sC^0\setminus\sC^{0,\ast}$, Lemma \ref
{legFirstFactorPositive} shows that the integrand equals $-\infty$
on a set of positive $F$-measure.
\end{pf}
\begin{Lemma}\label{legConcave}
The function $g$ is concave. If $\sC$ is convex, $g$ has at most one
maximum on $\sC\cap\sC^0$, modulo $\sN$.
\end{Lemma}
\begin{pf}
We first remark that the assertion is not trivial because $g$ need not
be strictly concave on $\sN^\bot$, for example,
the process $R_t=t (1,\ldots,1)^\top$ was not excluded.

Note that $g$ is of the form $g(y)= Hy+J(y)$, where
$Hy=L_{-}y^\top b^R + y^\top c^{RL}+ \int x' y^\top h(x) F^{R,L}$
is linear and $J(y)=\frac{(p-1)}{2}L_{-}y^\top c^R y +
I^{F^{R,L}}(y)$ is concave. We may assume that $h(x)=x1_{\{|x|\leq1\}}$.

Let $y_1,y_2\in\sC\cap\sC^0$ be such that $g(y_1)=g(y_2)=\sup g =:
g^*<\infty$, our aim is to show $y_1-y_2\in\sN$. By concavity,
$g^*=g((y_1+y_2)/2))=[g(y_1)+g(y_2)]/2$, which implies
$J((y_1+y_2)/2))=[J(y_1)+J(y_2)]/2$ due to the linearity of $H$. Using
the definition of $J$, this shows that $J$ is constant on the line
segment connecting $y_1$ and $y_2$. A first\vspace*{1.5pt} consequence is that the
difference
$y_1-y_2$ lies in the set $\{y\dvtx y^\top c^R=0, F^R\{x\dvtx y^\top
x\neq
0\}=0 \}$ and a second is that
$Hy_1=Hy_2$. It remains\vspace*{1pt} to show $(y_1-y_2)^\top b^R=0$ to have
$y_1-y_2\in\sN$.

Note that $F^R\{x\dvtx y^\top x\neq0\}=0$ implies $F^{R,L}\{x\dvtx
y^\top
h(x)\neq0\}=0$. Moreover,
$y^\top c^R=0$ implies $y^\top c^{RL}=0$ due to the absolute
continuity $\langle R^{c,i},L^c \rangle\ll\langle R^{c,i} \rangle$
which follows from the
Kunita--Watanabe inequality.
Therefore, the first consequence above implies $\int x' (y_1-y_2)^\top
h(x) F^{R,L}=0$ and $(y_1-\break y_2)^\top c^{RL}=0$, and now the second
consequence and the definition of $H$ yield
$0=H(y_1-y_2)=L_{-}(y_1-y_2)^\top b^R$. Thus $(y_1-y_2)^\top b^R=0$ as
$L_->0$ and the proof is complete.
\end{pf}

We can now move toward the main goal of this section.
Clearly we need some variant of the ``measurable maximum theorem''
(see, e.g., \cite{AliprantisBorder06}, 18.19;~\cite{KaratzasKardaras07}, Theorem~9.5;~\cite{Rockafellar76}, 2K).
We state a version that is tailored to our needs and has a simple
proof; the technique is used also
in Proposition \ref{prMeasSelectionContCase}.
\begin{Lemma}\label{lemeasMaxThm}
Let $\sD$ be a predictable set-valued process with nonempty compact
values in $2^{\R^d}$.
Let $f(y)=f(\omega,t,y)$ be a proper function on $\sD$ with values in
$\R\cup\{-\infty\}$
such that:
\begin{longlist}
\item$f(\varphi)$ is predictable whenever $\varphi$ is a $\sD$-valued
predictable process,
\item$y\mapsto f(y)$ is upper semicontinuous on $\sD$ for fixed
$(\omega,t)$.
\end{longlist}
Then there exists a $\sD$-valued predictable process $\pi$ such that
$f(\pi)=\max_{\sD} f$.
\end{Lemma}
\begin{pf}
We start with the Castaing representation (\cite{Rockafellar76}, 1B) %
of $\sD$:
there exist $\sD$-valued predictable processes $(\varphi_n)_{n\geq1}$
such that $\overline{\{\varphi_n\dvtx n\geq1\}}=\sD$ for each
$(\omega,t)$.
By (i), $f^*:=\max_n f(\varphi_n)$ is predictable, and
$f^*=\max_{\sD} f$ by (ii).
Fix $k\geq1$ and let $\Lambda_n:=\{f^*-f(\varphi_n)\leq1/k\}$,
$\Lambda^n:=\Lambda_n\setminus(\Lambda_1\cup\cdots\cup\Lambda_{n-1})$.
If we define $\pi^k:=\sum_{n} \varphi_n 1_{\Lambda^n}$, then $f^*-f(\pi
^k)\leq1/k$ and $\pi^k\in\sD$.

It remains to select a cluster point. By compactness, $(\pi^k)_{k\geq1}$
is bounded for each $(\omega,t)$, so there is a convergent subsequence
along ``random indices''~$\tau_k$.
More precisely,
there exists a strictly increasing sequence of integer-valued
predictable processes $\tau_k=\{\tau_k(\omega,t)\}$
and a predictable process $\pi^*$ such that
$\lim_k \pi^{\tau_k(\omega,t)}_t(\omega)= \pi^*_t(\omega)$ for all
$(\omega,t)$.
See, for example, the proof of F\"ollmer and Schied \cite
{FollmerSchied04}, Lemma 1.63.
We have $f^*=f(\pi^*)$ by (ii).
\end{pf}

Our random function $g$ satisfies property (i) of Lemma \ref
{lemeasMaxThm} because the characteristics are predictable (recall the
definition \cite{JacodShiryaev03}, II.1.6). We also note that the
intersection of closed predictable
processes is predictable
(\cite{Rockafellar76}, 1M). The sign of $p$ is
important as it switches the semicontinuity of $g$; we start with the
immediate case $p<0$ and denote $B_r(\R^d)=\{x\in\R^d\dvtx|x|\leq
r\}$.
\begin{pf*}{Proof of Lemma \ref{lemeasmaxsequence} for $p<0$}
In this case $g$ is u.s.c. on $\sC\cap\sC^0$ (Lem\-ma~\ref{lePropertiesOfI}).
Let $\sD(n):=\sC\cap\sC^0\cap B_n(\R^d)$. Lemma \ref{lemeasMaxThm}
yields a predictable process $\pi^n\in\argmax_{\sD(n)} g$
for each $n\geq1$, and clearly $\lim_n g(\pi^n)=\sup_{\sC\cap\sC
^0}g$. As $g(\pi^n)\geq g(0)=0$, we have $\pi^n\in\sC^{0,*}$ by
Lemma \ref{lePropertiesOfI}.
\end{pf*}

\subsection{Measurable maximizing sequence for $p\in(0,1)$}
\label{semeasMaxPpos}

Fix $p\in(0,1)$. Since the continuity properties of $g$ are not clear,
we will use an approximating sequence of continuous functions. (See
also Appendix \ref{setransformedModel}, where an alternative approach
is discussed and the continuity is clarified under an additional
assumption on $\sC$.)
We will approximate $g$ using L\'evy measures with enhanced
integrability, a method suggested by \cite{KaratzasKardaras07} in a
similar problem. This preserves monotonicity properties that will be
useful to pass to the limit.

All this is not necessary if $R$ is locally bounded, or more generally
if $F^{R,L}$ satisfies the following condition.
We start with fixed $(\omega,t)$.
\begin{Def}\label{depSuitable}
Let $F$ be a L\'evy measure on $\R^{d+1}$ which is equivalent to
$F^{R,L}$ and satisfies (\ref{eqFRLintegrates}).
(i) We say that $F$ is \textit{$p$-suitable} if
\[
\int(1+|x'|)(1+|x|)^p 1_{\{|x|>1\}} F(d(x,x'))<\infty.
\]

(ii) The \textit{$p$-suitable approximating sequence for $F$} is the
sequence $(F_n)_{n\geq1}$ of L\'evy measures defined by
$dF_n/dF=f_n$, where
\[
f_n(x)=1_{\{|x|\leq1\}}+e^{-|x|/n}1_{\{|x|>1\}}.
\]
\end{Def}

It is easy to see that each $F_n$ in (ii) shares the properties of $F$,
while in addition being $p$-suitable because $(1+|x|)^p e^{-|x|/n}$ is bounded.
As the sequence $f_n$ is increasing, monotone convergence shows that
$\int V \,dF_n \uparrow\int V \,dF$ for any measurable function $V\geq0$
on $\R^{d+1}$.
We denote by $g^F$ the function which is defined as in (\ref
{eqDefOfgRestated}) but with $F^{R,L}$ replaced by $F$.
\begin{Lemma}\label{lepSuitableCont}
If $F$ is $p$-suitable, $g^F$ is real-valued and continuous on $\sC^0$.
\end{Lemma}

\begin{pf}
Pick $y_n\to y$ in $\sC^0$. The only term in (\ref{eqDefOfgRestated})
for which continuity is not evident is the integral $I^F=I^F_\eps
+I^F_{>\eps}$,
where we choose $\eps$ as in Lem\-ma~\ref{lePropertiesOfI}. We have
$I^F_\eps(y_n)\to I^F_\eps(y)$ by that lemma. When $F$ is
$p$-suitable, the
continuity of $I^F_{>\eps}$ follows from the dominated convergence theorem.
\end{pf}
\begin{Remark}\label{rkdiamondDef}
Define the set
\[
(\sC\cap\sC^0)^{\diamond}:=\bigcup_{\eta\in[0,1)} \eta(\sC\cap
\sC^0).
\]
Its elements $y$ have the property that $1+y^\top x$ is
$F^R(dx)$-essentially \textup{boun\-ded away} from zero. Indeed, $y=\eta
y_0$ with $\eta\in[0,1)$ and $F^R\{y_0^\top x\geq-1\}=0$, therefore,
$1+y^\top x\geq1-\eta$, $F^R$-a.e.
In particular, $(\sC\cap\sC^0)^{\diamond}\subseteq\sC^{0,*}$. If
$\sC$ is star-shaped with respect to the origin, we also have $(\sC
\cap\sC^0)^{\diamond}\subseteq\sC$.
\end{Remark}

We introduce the compact-valued process $\sD(r):=\sC\cap\sC^0\cap
B_r(\R^d)$.
\begin{Lemma}\label{lediffUnderInt}
Let $F$ be $p$-suitable. Under \textup{(C3)}, $\argmax_{\sD(r)}
g^F\subseteq\sC^{0,*}$.

More generally, this holds whenever $F$ is a L\'evy measure
equivalent\break
to~$F^{R,L}$ satisfying (\ref{eqFRLintegrates})
and $g^F$ is finite-valued.
\end{Lemma}
\begin{pf}
Assume\vspace*{1pt} that $\cy\in\sC^0\setminus\sC^{0,*}$ is a maximum of $g^F$.
Let $\eta\in(\underline{\eta},1)$ be as in the definition of (C3) and
$y_0:=\eta\cy$. By Lemma \ref{lediffUnderIntForConcave}, the
directional derivative $D_{\cy,y_0}g$
can be calculated by differentiating under the integral sign. For the
integrand of $I^F$ we have
\[
D_{\cy,y_0}\{p^{-1}(1+y^\top x)^p - p^{-1} - y^\top h(x)\} =
(1-\eta)\{(1+ \cy^\top x)^{p-1} \cy^\top x - \cy^\top h(x)\}.
\]
But this is infinite on a set of positive measure as $\cy\in\sC
^0\setminus\sC^{0,*}$ means that $F\{\cy^\top x=-1\}>0$, contradicting
the last assertion of Lemma \ref{lediffUnderIntForConcave}.
\end{pf}

Let $F$ be a L\'evy measure on $\R^{d+1}$ which is equivalent to
$F^{R,L}$ and satisfies~(\ref{eqFRLintegrates}).
The following lemma is the crucial step in our argument.
\begin{Lemma}\label{leapproxgLastStep}
Let $(F_n)$ be the $p$-suitable approximating sequence for $F$ and fix $r>0$.
For each $n$, $\argmax_{\sD(r)} g^{F_n}\neq\varnothing$, and for any
$y_n^*\in\break\argmax_{\sD(r)} g^{F_n}$ it holds that
$
\limsup_n g^F(y^*_n) = \sup_{\sD(r)} g^{F}.
$
\end{Lemma}
\begin{pf}
We first show that
%
%
\begin{equation}\label{eqproofapproxgLastStep}
I^{F_n}(y)\to I^F(y)\qquad \mbox{for any }y\in\sC^0.
\end{equation}
Recall that
$
I^{F_n}(y)=\int(l+x') \{p^{-1}(1+y^\top x)^p - p^{-1} - y^\top
h(x)\}f_n(x) F(d(x,x')),
$
where $f_n$ is nonnegative and increasing in $n$.
As $f_n=1$ in a neighborhood of the origin, we need to consider only
$I^{F_n}_{>\eps}$ (for $\eps=1$, say). Its integrand is bounded below,
simultaneously for all $n$, by a negative constant times $(1+|x'|)$,
which is $F$-integrable on the relevant domain. As $(f_n)$ is
increasing, we can apply monotone convergence
on the set $\{(x,x')\dvtx p^{-1}(1+y^\top x)^p - p^{-1} - y^\top
h(x)\geq0 \}$ and dominated convergence on the complement to
deduce~(\ref{eqproofapproxgLastStep}).

Existence of $y^*_n\in\argmax_{\sD(r)} g^{F_n}$ is clear by
compactness of $\sD(r)$ and continuity of $g^{F_n}$ (Lemma \ref
{lepSuitableCont}). Let $y\in\sD(r)$ be arbitrary. By definition of~$y^*_n$ and (\ref{eqproofapproxgLastStep}),
\[
\limsup_n g^{F_n}(y^*_n)\geq\limsup_n g^{F_n}(y) = g^F(y).
\]
We show $\limsup_n g^{F}(y^*_n)\geq\limsup_n g^{F_n}(y^*_n)$. %
We can split the integral~$I^{F_n}(y)$ into a sum of three terms: the
integral over $\{|x|\leq1\}$ is the same as for
$I^F$, since $f_n=1$ on this set. We can assume that the cut-off $h$
vanishes outside $\{|x|\leq1\}$. The second term is then
\[
\int_{\{|x|>1\}} (l+x') p^{-1}(1+y^\top x)^p f_n \,dF,
\]
here the integrand is nonnegative and hence increasing in $n$, for
all $y$; and
the third term is
\[
\int_{\{|x|>1\}} (l+x') (-p^{-1}) f_n \,dF,
\]
which is decreasing in $n$ but converges to $\int_{\{|x|>1\}} (l+x')
(-p^{-1}) \,dF$. %
Thus we have that
\[
g^F(y_n^*)\geq g^{F_n}(y_n^*)- \eps_n
\]
with\vspace*{1pt} the sequence $\eps_n:=\int_{\{|x|>1\}} (l+x') (-p^{-1}) (f_n-1)
\,dF \downarrow0$.
Together, we conclude
$
\sup_{\sD(r)} g^{F} \geq\limsup_n g^{F}(y^*_n)\geq\limsup_n
g^{F_n}(y^*_n) \geq\sup_{\sD(r)} g^{F}.
$
\end{pf}
\begin{pf*}{Proof of Lemma \ref{lemeasmaxsequence} for $p\in(0,1)$}
Fix $r>0$. By Lemma \ref{lemeasMaxThm} we can find measurable
selectors $\pi^{n,r}$ for $\argmax_{\sD(r)} g^{F_n}$; that is,
$\pi^{n,r}_t(\omega)$ plays the role of $y_n^*$ in Lemma \ref
{leapproxgLastStep}. Taking $\pi^n:=\pi^{n,n}$ and noting that $\sD
(n)\uparrow\sC\cap\sC^0$, the
preceding Lemma \ref{leapproxgLastStep} shows that
$\pi
^{n}$ are $\sC\cap\sC^0$-valued predictable processes such that
$\limsup
_n g(\pi^n) = \sup_{\sC\cap\sC^0} g$ $P\otimes A$-a.e.
Lemma \ref{lediffUnderInt} shows that $\pi^n$ takes values in $\sC^{0,*}$.
\end{pf*}

\section{Parametrization by representative portfolios}
\label{setransformedModel}

This Appendix introduces an equivalent transformation of the model~%
$(R,\sC)$ with specific properties (Theorem \ref{thChangeCoords}). The
main idea is to substitute the given assets by wealth processes that
represent the investment opportunities of the model. While the result
is of independent interest, the main conclusion in our context is that
the approximation technique from Appendix \ref{semeasMaxPpos} for the
case $p\in(0,1)$ can be avoided, at least under slightly stronger
assumptions on $\sC$:
if the utility maximization problem is finite, the corresponding L\'evy
measure in the transformed model is $p$-suitable (cf. Definition \ref
{depSuitable}) and hence the corresponding function $g$ is continuous.
This is not only an alternative argument to prove Lemma \ref
{lemeasmaxsequence}. In applications, continuity can be useful to
construct a maximizer for $g$ (rather than a maximizing sequence) if
one does not know a priori that there exists an optimal strategy.
A~static version of our construction was carried out for the case of L\'
evy processes in~\cite{Nutz09c}, Section 4.

In this Appendix we use the following assumptions on the
set-valued process $\sC$ of constraints:
\begin{longlist}[(C4)]
\item[(C1)] $\sC$ is predictable.
\item[(C2)] $\sC$ is closed.
\item[(C4)] $\sC$ is star-shaped with respect to the origin: $\eta
\sC
\subseteq\sC$ for all $\eta\in[0,1]$.
\end{longlist}

Since we already obtained a proof of Lemma \ref{lemeasmaxsequence}, we
do not strive for minimal conditions here.
Clearly (C4) implies condition (C3) from Section~\ref
{seConstraintsDegeneracies}, but its main implication is that we can
select a bounded (hence $R$-integrable) process in the subsequent
lemma. The following result is the construction of the $j$th \textit
{representative portfolio}, a portfolio with the property that it
invests in the $j$th asset \textit{whenever} this is feasible.
\begin{Lemma}\label{lerepresentativePortf}
Fix $1\leq j\leq d$ and let $H^j=\{x\in\R^d\dvtx x^j\neq0\}$. There
exists a bounded predictable $\sC\cap\sC^{0,*}$-valued process $\phi
$ satisfying
\[
\{\phi^j= 0\} = \{ \sC\cap\sC^{0,*}\cap H^j = \varnothing\}.
\]
\end{Lemma}
\begin{pf}
Let $B_1=B_1(\R^d)$ be the closed unit ball and $H:=H^j$. Condition
(C4) implies $\{ \sC\cap\sC^{0,*}\cap H = \varnothing\}=\{
\sC\cap B_1\cap\sC^{0,*}\cap H = \varnothing\}$, hence, we may
substitute $\sC$ by $\sC\cap B_1$.
Define the closed sets $H_k=\{x\in\R^d\dvtx|x^j|\geq k^{-1}\}$ for
\mbox{$k\geq1$}, then $\bigcup_k H_k=H$.
Moreover, let $\sD_k=\sC\cap\sC^0\cap H_k$. This is a~compact-valued
predictable process, so there exists a predictable process~$\phi_k$
such that $\phi_k\in\sD_k$ (hence $\phi_k^j\neq0$) on the set
$\Lambda
_k:=\{\sD_k\neq\varnothing\}$ and $\phi_k=0$ on the complement.
Define $\Lambda^k:=\Lambda_k\setminus(\Lambda_1\cup\cdots\cup
\Lambda
_{k-1})$ and
$\phi':=\sum_{k} \phi_k 1_{\Lambda^k}$. Then $|\phi'|\leq1$ and
$\{\phi'^j= 0\} = \{ \sC\cap\sC^{0}\cap H = \varnothing\}=
\{ \sC\cap\sC^{0,*}\cap H = \varnothing\}$; the second equality uses
(C4) and Remark \ref{rkdiamondDef}. These two facts also show that
$\phi
:=\frac{1}{2}\phi'$ has the same property while in addition being
$\sC
\cap\sC^{0,*}$-valued.
\end{pf}
\begin{Remark}
The previous proof also applies if instead of \textup{(C4)}, for example, the
diameter of $\sC$ is uniformly bounded and
$\sC^0=\sC^{0,*}$.
\end{Remark}

If $\Phi$ is a $d\times d$-matrix with columns $\phi_1,\ldots,\phi
_d\in
L(R)$, the matrix stochastic integral
$\tR=\Phi\sint R$ is the $\R^d$-valued process given by $\tR^j=\phi
_j\sint R$. Moreover, if $\psi\in L(\Phi\sint R)$
is $\R^d$-valued, then $\Phi\psi\in L(R)$ and
%
%
\begin{equation}\label{eqassocIntegral}
\psi\sint(\Phi\sint R) = (\Phi\psi)\sint R.
\end{equation}
If $\sD$ is a set-valued process which is predictable, closed and
contains the origin, then the pre-image $\Phi^{-1}\sD$ shares these
properties; cf. \cite{Rockafellar76}, 1Q. Convexity and star-shape are
also preserved.

We obtain\vspace*{1pt} the following model if we sequentially replace the given
assets by representative portfolios;
here $e_j$ denotes the $j$th unit vector in $\R^d$ for $1\leq j\leq d$
(i.e., $e^i_j=\delta_{ij}$).
\begin{Thm}\label{thChangeCoords}
There exists a predictable $\R^{d\times d}$-valued uniformly boun\-ded
process $\Phi$ such that
the financial market model with returns
\[
\tR:=\Phi\sint R
\]
and constraints $\tsC:=\Phi^{-1} \sC$ has the following properties:
for all $1\leq j\leq d$,
\begin{longlist}
\item$\Delta\tR^j> -1$ (positive prices),\vspace*{2pt}
\item$e_j\in\tsC\cap\tsC^{\,0,*}$, where $\tsC^{\,0,*}=\Phi^{-1}\sC^{0,*}$
(entire wealth can be invested in each asset),
\item the model $(\tR,\tsC)$ admits the same wealth processes as
$(R,\sC)$.
\end{longlist}
\end{Thm}
\begin{pf}
We treat\vspace*{1pt} the components one by one. Let $j=1$ and let $\phi=\phi(1)$
be as in Lemma \ref{lerepresentativePortf}. We replace the first asset
$R^1$ by the process $\phi\sint R$, or equivalently, we replace $R$ by
$\Phi\sint R$, where $\Phi=\Phi(1)$ is the $d\times d$-matrix
\[
\Phi= \pmatrix{
\phi^1 & & & \cr
\phi^2 & 1 & & \cr
\vdots& & \ddots& \cr
\phi^d & & & 1}.
\]
The new natural constraints are $\Phi^{-1} \sC^0$ and we replace $\sC$
by $\Phi^{-1}\sC$.
Note that $e_1\in\Phi^{-1} (\sC\cap\sC^{0,*})$ because $\Phi
e_1=\phi
\in\sC\cap\sC^{0,*}$ by construction.

We show that for every $\sC\cap\sC^{0,*}$-valued process $\pi\in L(R)$
there exists~$\psi$ predictable such that
$\Phi\psi=\pi$. In view of (\ref{eqassocIntegral}), this will imply
that the new model admits the same wealth processes as the old one.
On the set $\{\phi^1\neq0\}=\{\Phi\mbox{ is invertible}\}$ we take
$\psi=\Phi^{-1}\pi$ and on the complement we
choose $\psi^1\equiv0$ and $\psi^j=\pi^j$ for $j\geq2$; this is the
same as inverting $\Phi$ on its image.
Note that $\{\phi^1=0\}\subseteq\{\pi^1=0\}$ by the choice of $\phi$.

We proceed with the second component of the new model in the same way,
and then continue until the last one.
We obtain matrices $\Phi(j)$ for $1\leq j\leq d$ and set $\hat{\Phi
}=\Phi(1) \cdots\Phi(d)$. Then
$\hat{\Phi}$ has the required properties. Indeed,
the construction and $\Phi(i) e_j=e_j$ for $i\neq j$ imply $e_j \in
\hat{\Phi}^{-1} (\sC\cap\sC^{0,*})$.
This is (ii), and (i) is a consequence of (ii).
\end{pf}

Coming back to the utility maximization problem, note that property~(iii) implies that the value functions and the opportunity
processes for the models~$(R,\sC)$ and $(\tR,\tsC)$ coincide up to
evanescence; we identify them in the sequel. Furthermore,
if $\tg$ denotes the analogue of $g$ in the model $(\tR,\tsC)$,
cf.~(\ref{eqDefOfgRestated}), we have the relation
\[
\tg(y)=g(\Phi y),\qquad y\in\tsC^{\,0}.
\]
Finding a maximizer for $\tg$ is equivalent to finding one for $g$ and
if $(\tpi,\kappa)$
is an optimal strategy for $(\tR,\tsC)$, then $(\Phi\tpi,\kappa)$ is
optimal for $(R,\sC)$.
In fact, most properties of interest carry over from $(R,\sC)$ to
$(\tR
,\tsC)$,
in particular any no-arbitrage property that is defined via the set of
admissible (positive) wealth processes.
\begin{Remark}
A classical no-arbitrage condition defined in a slightly different way
is that there exist a probability
measure $Q\approx P$ under which~$\cE(R)$ is a $\sigma$-martingale;
cf. Delbaen and Schachermayer \cite{DelbaenSchachermayer98}. In this case,
$\cE(\tR)$ is even a local martingale under $Q$, as it is a $\sigma
$-martingale with positive components.
\end{Remark}

Property (ii) from Theorem \ref{thChangeCoords} is useful to
apply the following result.

\begin{Lemma}\label{leRecoverpSuitable}
Let $p\in(0,1)$ and assume $e_j\in\sC\cap\sC^{0,*}$ for $1\leq
j\leq d$. Then $u(x_0)<\infty$ implies
that $F^{R,L}$ is $p$-suitable. If, in addition, there exists a
constant $k_1$ such that
$D\geq k_1>0$, it follows that $\int_{\{|x|>1\}} |x|^p F^R(dx)<\infty$.
\end{Lemma}

\begin{pf}
As $p>0$ and $u(x_0)<\infty$, $L$ is well defined and $L,L_->0$, by
Section \ref{seoppProc}.
No further properties were used to establish Lemma \ref
{leCompensatorJointChar}, whose formula shows that $g(\pi)$ is finite
$P\otimes A$-a.e. for all $\pi\in\cA=\cA^{fE}$. In particular, from
the definition of $g$, it follows that
$\int(L_-+x') \{p^{-1}(1+\pi^\top x)^p - p^{-1} - \pi^\top
h(x)\} F^{R,L}(d(x,x'))$ is finite. If $D\geq k_1$, \cite
{Nutz09a}, Lemma 3.5, shows that $L\geq k_1$, hence, $L_-+x'\geq k_1$
$F^L(dx')$-a.e. and
$\int\{p^{-1}(1+\pi^\top x)^p - p^{-1} - \pi^\top h(x)\}
F^{R}(dx)<\infty$.
We choose $\pi=e_j$ (and $\kappa$ arbitrary) for $1\leq j\leq d$ to
deduce the result.
\end{pf}

In general, the condition $u(x_0)<\infty$ does not imply any properties
of $R$; for instance, in the trivial cases $\sC=\{0\}$ or $\sC
^{0,*}=\{
0\}$.
The transformation changes the geometry of $\sC$ and $\sC^{0,*}$ such
that Theorem \ref{thChangeCoords}(ii) holds, and then the
situation is different.
\begin{Cor}
Let $p\in(0,1)$ and $u(x_0)<\infty$. In the model $(\tR,\tsC)$ of
Theorem \ref{thChangeCoords}, $F^{\tR,L}$ is $p$-suitable
and hence, $\tg$ is continuous.
\end{Cor}

Therefore, to prove Lemma \ref{lemeasmaxsequence} under (C4), we may
substitute $(R,\sC)$ by~$(\tR,\tsC)$ and avoid the use of $p$-suitable
approximating sequences. In some cases, Lem\-ma~\ref{leRecoverpSuitable}
applies\vspace*{1pt} directly in $(R,\sC)$.
In particular, if the asset prices are strictly positive ($\Delta
R^j>-1$ for $1\leq j\leq d$), then the positive orthant of~$\R^d$ is
contained in $\sC^{0,*}$ and the condition of Lemma \ref
{leRecoverpSuitable} is satisfied as soon as $e_j\in\sC$ for $1\leq
j\leq d$.

\section{Omitted calculation}

This Appendix contains a calculation which was omitted in the proof of
Proposition \ref{prsupermartAndGradient}.\vadjust{\goodbreak}
\begin{Lemma}\label{leomittedCalc}
Let $(\ell,\cpi,\ckappa)$ be a solution of the Bellman equation,
$(\pi,\kappa)\in\cA$, $X:=X(\pi,\kappa)$ and $\cX:=X(\cpi
,\ckappa)$.
Define $\bar{R}= R - (x-h(x))\ast\mu^R$ as well as $\bar{\pi
}:=(p-1)\cpi+\pi$ and $\bar{\kappa}:=(p-1)\ckappa+\kappa$. Then
$\xi
:=\ell\cX^{p-1}X$ satisfies
\begin{eqnarray*}
&&(\cX_-^{p-1} X_-)^{-1}\sint\xi\\
&&\qquad=\ell-\ell_0 + \ell_-\bar{\pi}\sint\bar{R} - \ell_- \bar
{\kappa}\sint
\mu\\
&&\qquad\quad{}+ \ell_-(p-1)\biggl(\frac{p-2}{2}\cpi+\pi\biggr)^\top c^R \cpi\sint A
+ \bar{\pi}^\top c^{R\ell}\sint A
+ \bar{\pi}^\top x'h(x) \ast\mu^{R,\ell}\\
&&\qquad\quad{}+ (\ell_-+x')\{(1+\cpi^\top x)^{p-1}(1+\pi^\top x) - 1
- \bar{\pi}^\top h(x)\}\ast\mu^{R,\ell}.
\end{eqnarray*}
\end{Lemma}
\begin{pf}
We may assume $x_0=1$. This calculation is similar to the one in the
proof of Lemma \ref{leCompensatorJointChar}
and, therefore, we shall be brief. By It\^o's formula we have $\cX
^{p-1}=\cE(\zeta)$ for
\begin{eqnarray*}
\zeta &=& (p-1)(\cpi\sint R - \ckappa\sint\mu) + \frac
{(p-1)(p-2)}{2}\cpi^\top c^R\cpi\sint A \\
&&{} + \{(1+\cpi^\top x)^{p-1}-1-(p-1)\cpi^\top x\}\ast\mu^R.
\end{eqnarray*}
Thus $\cX^{p-1} X=\cE(\zeta+ \pi\sint R - \kappa\sint\mu+
[\zeta,\pi\sint R])=:\cE(\Psi)$ with
\begin{eqnarray*}
[R,\zeta] &=& [R^c,\zeta^c]+ \sum\Delta R \Delta\zeta\\
&=& (p-1)c^R \cpi\sint A + (p-1)\cpi^\top x x\ast\mu^R\\
&&{} +x\{(1+\cpi^\top x)^{p-1}-1-\cpi^\top x\}\ast\mu^R
\end{eqnarray*}
and recombining the terms yields
\begin{eqnarray*}
\Psi &=& \bar{\pi} \sint R - \bar{\kappa} \sint\mu+ (p-1)\biggl(\frac
{p-2}{2}\cpi+\pi\biggr)^\top c^R \cpi\sint A\\
&&{} + \{(1+\cpi^\top x)^{p-1}(1+\pi^\top x) - 1 - \bar{\pi}^\top
x\} \ast\mu^{R}.
\end{eqnarray*}
Then $(\cX_-^{p-1} X_-)^{-1}\sint\xi= \ell-\ell_0 +\ell
_-\sint\Psi+ [\ell,\Psi]$, where
\begin{eqnarray*}
[\ell,\Psi] & = & [\ell^c,\Psi^c]+ \sum\Delta\ell\Delta\Psi\\
& = & \bar{\pi}^\top c^{R\ell}\sint A + \bar{\pi}^\top x'x \ast\mu
^{R,\ell}\\
&&{} + x'\{(1+\cpi^\top x)^{p-1}(1+\pi^\top x) - 1 - \bar{\pi}^\top
x\} \ast\mu^{R,\ell}.
\end{eqnarray*}
We arrive at
\begin{eqnarray*}
&&(\cX_-^{p-1} X_-)^{-1}\sint\xi\\
&&\qquad=\ell-\ell_0 + \ell_-\bar{\pi}\sint R - \ell_- \bar{\kappa
}\sint\mu\\
&&\qquad\quad{}+
\ell_-(p-1)\biggl(\frac{p-2}{2}\cpi+\pi\biggr)^\top c^R \cpi\sint A + \bar
{\pi}^\top c^{R\ell}\sint A + \bar{\pi}^\top x'x \ast\mu^{R,\ell}\\
&&\qquad\quad{} + (\ell_-+x')\{(1+\cpi
^\top x)^{p-1}(1+\pi^\top x) - 1 - \bar{\pi}^\top x\} \ast\mu
^{R,\ell}.
\end{eqnarray*}
The result follows by writing $x=h(x)+x-h(x)$.
\end{pf}
\end{appendix}

\section*{Acknowledgments}
The author thanks Christoph Czichowsky for fruitful discussions, and
Martin Schweizer, Nicholas Westray and an anonymous referee for
comments on an earlier version of the manuscript.


%

%
\printaddresses


\begin{thebibliography}{30}

\bibitem{AliprantisBorder06}
%
\begin{bbook}[mr]
\bauthor{\bsnm{Aliprantis},~\bfnm{Charalambos~D.}\binits{C.~D.}}
\AND
\bauthor{\bsnm{Border},~\bfnm{Kim~C.}\binits{K.~C.}}
(\byear{2006}).
\btitle{Infinite Dimensional Analysis: A~Hitchhiker's Guide}, \bedition{3rd} ed.
\bpublisher{Springer}, \baddress{Berlin}.
\bid{mr={2378491}}
\end{bbook}
%
\endbibitem

\bibitem{AnselStricker03}
%
\begin{bincollection}[mr]
\bauthor{\bsnm{Ansel},~\bfnm{J.~P.}\binits{J.~P.}} \AND
\bauthor{\bsnm{Stricker},~\bfnm{C.}\binits{C.}}
(\byear{1993}).
\btitle{D\'ecomposition de {K}unita--{W}atanabe}.
In \bbooktitle{S\'eminaire de {P}robabilit\'es, {XXVII}}.
\bseries{Lecture Notes in Math.}
\bvolume{1557}
\bpages{30--32}.
\bpublisher{Springer}, \baddress{Berlin}.
\bid{doi={10.1007/BFb0087960}, mr={1308549}}
\end{bincollection}
%
\endbibitem

\bibitem{BarndorffNielsenShephard01}
%
\begin{barticle}[mr]
\bauthor{\bsnm{Barndorff-Nielsen},~\bfnm{Ole~E.}\binits{O.~E.}}
\AND
\bauthor{\bsnm{Shephard},~\bfnm{Neil}\binits{N.}}
(\byear{2001}).
\btitle{Non-{G}aussian {O}rnstein--{U}hlenbeck-based models and some
of their
uses in financial economics}.
\bjournal{J.~R.~Stat. Soc. Ser. B Stat. Methodol.}
\bvolume{63}
\bpages{167--241}.
\bid{doi={10.1111/1467-9868.00282}, issn={1369-7412}, mr={1841412}}
\end{barticle}
%
\endbibitem

\bibitem{BriandHu08}
%
\begin{barticle}[mr]
\bauthor{\bsnm{Briand},~\bfnm{Philippe}\binits{P.}} \AND
\bauthor{\bsnm{Hu},~\bfnm{Ying}\binits{Y.}}
(\byear{2008}).
\btitle{Quadratic {BSDE}s with convex generators and unbounded terminal
conditions}.
\bjournal{Probab. Theory Related Fields}
\bvolume{141}
\bpages{543--567}.
\bid{doi={10.1007/s00440-007-0093-y}, issn={0178-8051}, mr={2391164}}
\end{barticle}
%
\endbibitem

\bibitem{CernyKallsen07}
%
\begin{barticle}[mr]
\bauthor{\bsnm{{\v{C}}ern{\'y}},~\bfnm{Ale{\v{s}}}\binits{A.}}
\AND
\bauthor{\bsnm{Kallsen},~\bfnm{Jan}\binits{J.}}
(\byear{2007}).
\btitle{On the structure of general mean--variance hedging strategies}.
\bjournal{Ann. Probab.}
\bvolume{35}
\bpages{1479--1531}.
\bid{doi={10.1214/009117906000000872}, issn={0091-1798}, mr={2330978}}
\end{barticle}
%
\endbibitem

\bibitem{ChoulliStricker09}
%
\begin{barticle}[mr]
\bauthor{\bsnm{Choulli},~\bfnm{Tahir}\binits{T.}} \AND
\bauthor{\bsnm{Stricker},~\bfnm{Christophe}\binits{C.}}
(\byear{2009}).
\btitle{Comparing the minimal {H}ellinger martingale measure of order
{$q$} to
the {$q$}-optimal martingale measure}.
\bjournal{Stochastic Process. Appl.}
\bvolume{119}
\bpages{1368--1385}.
\bid{doi={10.1016/j.spa.2008.07.002}, issn={0304-4149}, mr={2508578}}
\end{barticle}
%
\endbibitem

\bibitem{CzichowskySchweizer09b}
%
\begin{bmisc}[auto:STB|2011-03-03|12:04:44]
\bauthor{\bsnm{Czichowsky},~\bfnm{C.}\binits{C.}} \AND
\bauthor{\bsnm{Schweizer},~\bfnm{M.}\binits{M.}}
(\byear{2011}).
\bhowpublished{On the Markowitz problem under convex cone constraints.
Unpublished manuscript}.
\end{bmisc}
%
\endbibitem

\bibitem{CzichowskySchweizer09a}
%
\begin{bincollection}[auto:STB|2011-03-03|12:04:44]
\bauthor{\bsnm{Czichowsky},~\bfnm{C.}\binits{C.}} \AND
\bauthor{\bsnm{Schweizer},~\bfnm{M.}\binits{M.}}
(\byear{2011}).
\btitle{Closedness in the semimartingale topology for spaces of stochastic
integrals with constrained integrands}.
In \bbooktitle{S\'eminaire de Probabilit\'es XLIII}.
\bseries{Lecture Notes in Math.}
\bvolume{2006}
\bpages{413--436}.
\bpublisher{Springer}, \baddress{Berlin}.
\end{bincollection}
%
\endbibitem

\bibitem{DelbaenSchachermayer98}
%
\begin{barticle}[mr]
\bauthor{\bsnm{Delbaen},~\bfnm{F.}\binits{F.}} \AND
\bauthor{\bsnm{Schachermayer},~\bfnm{W.}\binits{W.}}
(\byear{1998}).
\btitle{The fundamental theorem of asset pricing for unbounded stochastic
processes}.
\bjournal{Math. Ann.}
\bvolume{312}
\bpages{215--250}.
\bid{doi={10.1007/s002080050220}, issn={0025-5831}, mr={1671792}}
\end{barticle}
%
\endbibitem

\bibitem{FollmerSchied04}
%
\begin{bbook}[mr]
\bauthor{\bsnm{F{\"o}llmer},~\bfnm{Hans}\binits{H.}} \AND
\bauthor{\bsnm{Schied},~\bfnm{Alexander}\binits{A.}}
(\byear{2004}).
\btitle{Stochastic Finance: An Introduction in Discrete Time},
\bedition{2nd} ed.
\bseries{de Gruyter Studies in Mathematics}
\bvolume{27}.
\bpublisher{de Gruyter}, \baddress{Berlin}.
\bid{doi={10.1515/9783110212075}, mr={2169807}}
\end{bbook}
%
\endbibitem

\bibitem{GollKallsen03}
%
\begin{barticle}[mr]
\bauthor{\bsnm{Goll},~\bfnm{Thomas}\binits{T.}} \AND
\bauthor{\bsnm{Kallsen},~\bfnm{Jan}\binits{J.}}
(\byear{2003}).
\btitle{A complete explicit solution to the log-optimal portfolio problem}.
\bjournal{Ann. Appl. Probab.}
\bvolume{13}
\bpages{774--799}.
\bid{doi={10.1214/aoap/1050689603}, issn={1050-5164}, mr={1970286}}
\end{barticle}
%
\endbibitem

\bibitem{GollRuschendorf01}
%
\begin{barticle}[mr]
\bauthor{\bsnm{Goll},~\bfnm{Thomas}\binits{T.}} \AND
\bauthor{\bsnm{R{\"u}schendorf},~\bfnm{Ludger}\binits{L.}}
(\byear{2001}).
\btitle{Minimax and minimal distance martingale measures and their relationship
to portfolio optimization}.
\bjournal{Finance Stoch.}
\bvolume{5}
\bpages{557--581}.
\bid{doi={10.1007/s007800100052}, issn={0949-2984}, mr={1862002}}
\end{barticle}
%
\endbibitem

\bibitem{Grandits00}
%
\begin{barticle}[mr]
\bauthor{\bsnm{Grandits},~\bfnm{P.}\binits{P.}}
(\byear{2000}).
\btitle{On martingale measures for stochastic processes with independent
increments}.
\bjournal{Theory Probab. Appl.}
\bvolume{44}
\bpages{39--50}.
\bid{doi={10.1137/S0040585X97977355}, issn={0040-361X}, mr={1751190}}
\bptnote{check year}%
\end{barticle}
%
\endbibitem

\bibitem{HuImkellerMuller05}
%
\begin{barticle}[mr]
\bauthor{\bsnm{Hu},~\bfnm{Ying}\binits{Y.}},
\bauthor{\bsnm{Imkeller},~\bfnm{Peter}\binits{P.}} \AND
\bauthor{\bsnm{M{\"u}ller},~\bfnm{Matthias}\binits{M.}}
(\byear{2005}).
\btitle{Utility maximization in incomplete markets}.
\bjournal{Ann. Appl. Probab.}
\bvolume{15}
\bpages{1691--1712}.
\bid{doi={10.1214/105051605000000188}, issn={1050-5164}, mr={2152241}}
\end{barticle}
%
\endbibitem

\bibitem{JacodShiryaev03}
%
\begin{bbook}[mr]
\bauthor{\bsnm{Jacod},~\bfnm{Jean}\binits{J.}} \AND
\bauthor{\bsnm{Shiryaev},~\bfnm{Albert~N.}\binits{A.~N.}}
(\byear{2003}).
\btitle{Limit Theorems for Stochastic Processes},
\bedition{2nd} ed.
\bseries{Grundlehren der Mathematischen Wissenschaften [Fundamental Principles
of Mathematical Sciences]}
\bvolume{288}.
\bpublisher{Springer}, \baddress{Berlin}.
\bid{mr={1943877}}
\end{bbook}
%
\endbibitem

\bibitem{JeanblancKloppelMiyahara07}
%
\begin{barticle}[mr]
\bauthor{\bsnm{Jeanblanc},~\bfnm{Monique}\binits{M.}},
\bauthor{\bsnm{Kl{\"o}ppel},~\bfnm{Susanne}\binits{S.}} \AND
\bauthor{\bsnm{Miyahara},~\bfnm{Yoshio}\binits{Y.}}
(\byear{2007}).
\btitle{Minimal {$f\sp q$}-martingale measures of exponential {L}\'evy
processes}.
\bjournal{Ann. Appl. Probab.}
\bvolume{17}
\bpages{1615--1638}.
\bid{doi={10.1214/07-AAP439}, issn={1050-5164}, mr={2358636}}
\end{barticle}
%
\endbibitem

\bibitem{Kallsen04}
%
\begin{barticle}[mr]
\bauthor{\bsnm{Kallsen},~\bfnm{J.}\binits{J.}}
(\byear{2004}).
\btitle{{$\sigma$}-localization and {$\sigma$}-martingales}.
\bjournal{Theory Probab. Appl.}
\bvolume{48}
\bpages{152--163}.
\bid{doi={10.1137/S0040585X980312}, issn={0040-361X}, mr={2013413}}
\bptnote{check year}%
\end{barticle}
%
\endbibitem

\bibitem{KallsenMuhleKarbe08}
%
\begin{barticle}[mr]
\bauthor{\bsnm{Kallsen},~\bfnm{Jan}\binits{J.}} \AND
\bauthor{\bsnm{Muhle-Karbe},~\bfnm{Johannes}\binits{J.}}
(\byear{2010}).
\btitle{Utility maximization in affine stochastic volatility models}.
\bjournal{Int. J. Theor. Appl. Finance}
\bvolume{13}
\bpages{459--477}.
\bid{doi={10.1142/S0219024910005851}, issn={0219-0249}, mr={2660149}}
\end{barticle}
%
\endbibitem

\bibitem{KaratzasKardaras07}
%
\begin{barticle}[mr]
\bauthor{\bsnm{Karatzas},~\bfnm{Ioannis}\binits{I.}} \AND
\bauthor{\bsnm{Kardaras},~\bfnm{Constantinos}\binits{C.}}
(\byear{2007}).
\btitle{The num\'eraire portfolio in semimartingale financial models}.
\bjournal{Finance Stoch.}
\bvolume{11}
\bpages{447--493}.
\bid{doi={10.1007/s00780-007-0047-3}, issn={0949-2984}, mr={2335830}}
\end{barticle}
%
\endbibitem

\bibitem{KaratzasZitkovic03}
%
\begin{barticle}[mr]
\bauthor{\bsnm{Karatzas},~\bfnm{Ioannis}\binits{I.}} \AND
\bauthor{\bsnm{{\v{Z}}itkovi{\'c}},~\bfnm{Gordan}\binits{G.}}
(\byear{2003}).
\btitle{Optimal consumption from investment and random endowment in incomplete
semimartingale markets}.
\bjournal{Ann. Probab.}
\bvolume{31}
\bpages{1821--1858}.
\bid{doi={10.1214/aop/1068646367}, issn={0091-1798}, mr={2016601}}
\end{barticle}
%
\endbibitem

\bibitem{KramkovSchachermayer99}
%
\begin{barticle}[mr]
\bauthor{\bsnm{Kramkov},~\bfnm{D.}\binits{D.}} \AND
\bauthor{\bsnm{Schachermayer},~\bfnm{W.}\binits{W.}}
(\byear{1999}).
\btitle{The asymptotic elasticity of utility functions and optimal investment
in incomplete markets}.
\bjournal{Ann. Appl. Probab.}
\bvolume{9}
\bpages{904--950}.
\bid{doi={10.1214/aoap/1029962818}, issn={1050-5164}, mr={1722287}}
\end{barticle}
%
\endbibitem

\bibitem{KramkovSchachermayer03}
%
\begin{barticle}[mr]
\bauthor{\bsnm{Kramkov},~\bfnm{D.}\binits{D.}} \AND
\bauthor{\bsnm{Schachermayer},~\bfnm{W.}\binits{W.}}
(\byear{2003}).
\btitle{Necessary and sufficient conditions in the problem of optimal
investment in incomplete markets}.
\bjournal{Ann. Appl. Probab.}
\bvolume{13}
\bpages{1504--1516}.
\bid{doi={10.1214/aoap/1069786508}, issn={1050-5164}, mr={2023886}}
\end{barticle}
%
\endbibitem

\bibitem{ManiaTevzadze03}
%
\begin{barticle}[mr]
\bauthor{\bsnm{Mania},~\bfnm{M.}\binits{M.}} \AND
\bauthor{\bsnm{Tevzadze},~\bfnm{R.}\binits{R.}}
(\byear{2003}).
\btitle{A unified characterization of {$q$}-optimal and minimal entropy
martingale measures by semimartingale backward equations}.
\bjournal{Georgian Math. J.}
\bvolume{10}
\bpages{289--310}.
\bid{issn={1072-947X}, mr={2009977}}
\end{barticle}
%
\endbibitem

\bibitem{MuhleKarbe09}
%
\begin{bmisc}[auto:STB|2011-03-03|12:04:44]
\bauthor{\bsnm{Muhle-Karbe},~\bfnm{J.}\binits{J.}}
(\byear{2009}).
\bhowpublished{On utility-based investment, pricing and hedging in
incomplete markets. Ph.D. thesis, TU M\"unchen}.
\end{bmisc}
%
\endbibitem

\bibitem{Nutz09a}
%
\begin{barticle}[mr]
\bauthor{\bsnm{Nutz},~\bfnm{Marcel}\binits{M.}}
(\byear{2010}).
\btitle{The opportunity process for optimal consumption and investment with
power utility}.
\bjournal{Math. Financ. Econ.}
\bvolume{3}
\bpages{139--159}.
\bid{doi={10.1007/s11579-010-0031-0}, issn={1862-9679}, mr={2721695}}
\end{barticle}
%
\endbibitem

\bibitem{Nutz09c}
%
\begin{bmisc}[auto:STB|2011-03-03|12:04:44]
\bauthor{\bsnm{Nutz},~\bfnm{M.}\binits{M.}}
(\byear{2010}).
\bhowpublished{Power utility maximization in constrained exponential
L\'evy
models. \textit{Math. Finance}. To appear.
\href{http://dx.doi.org/10.1007/s00440-010-0334-3}{DOI:10.1007/s00440-010-0334-3}.}
\end{bmisc}
%
\endbibitem

\bibitem{Nutz09d}
%
\begin{bmisc}[auto:STB|2011-03-03|12:04:44]
\bauthor{\bsnm{Nutz},~\bfnm{M.}\binits{M.}}
(\byear{2011}).
\bhowpublished{Risk aversion asymptotics for power utility
maximization. \textit{Probab. Theory Related Fields}.
To appear.}
\end{bmisc}
%
\endbibitem

\bibitem{Rockafellar76}
%
\begin{bincollection}[mr]
\bauthor{\bsnm{Rockafellar},~\bfnm{R.~Tyrrell}\binits{R.~T.}}
(\byear{1976}).
\btitle{Integral functionals, normal integrands and measurable selections}.
In \bbooktitle{Nonlinear Operators and the Calculus of Variations ({S}ummer
{S}chool, {U}niv. {L}ibre {B}ruxelles, {B}russels, 1975)}.
\bseries{Lecture Notes in Math.}
\bvolume{543}
\bpages{157--207}.
\bpublisher{Springer}, \baddress{Berlin}.
\bid{mr={0512209}}
\end{bincollection}
%
\endbibitem

\bibitem{Schweizer95b}
%
\begin{barticle}[mr]
\bauthor{\bsnm{Schweizer},~\bfnm{Martin}\binits{M.}}
(\byear{1995}).
\btitle{On the minimal martingale measure and the {F}\"ollmer--{S}chweizer
decomposition}.
\bjournal{Stoch. Anal. Appl.}
\bvolume{13}
\bpages{573--599}.
\bid{doi={10.1080/07362999508809418}, issn={0736-2994}, mr={1353193}}
\end{barticle}
%
\endbibitem

\end{thebibliography}
\end{document}